\documentclass[a4paper,fleqn,usenatbib]{mnras}

\usepackage{float}
\usepackage[T1]{fontenc}
\usepackage{aecompl}
\usepackage{graphicx,color,bm}
\usepackage{latexsym}
\usepackage{epsfig}
\usepackage{amssymb}
\usepackage{amsmath}
\usepackage{wasysym}
\usepackage{pstricks}
\usepackage{enumitem}

\def\etal{{\frenchspacing\it et al.}}
\def\ie{{\frenchspacing\it i.e.}}
\def\eg{{\frenchspacing\it e.g.}}

\def\be{\begin{equation}}
\def\ee{\end{equation}}
\def\ba{\begin{eqnarray}}
\def\ea{\end{eqnarray}}

\def\nn{\nonumber}

\def\vunits{{\,h^3{\rm Mpc}^{-3}}}
\def\sqdeg{{\,{\rm deg}^2}}

\def\LaTeX{L\kern-.36em\raise.3ex\hbox{a}\kern-.15em
    T\kern-.1667em\lower.7ex\hbox{E}\kern-.125emX}

\begin{document}

\voffset-1.25cm
\title[Science forecast for the eBOSS survey]{The extended Baryon Oscillation Spectroscopic Survey (eBOSS): a cosmological forecast}
\author[Zhao \etal]{
\parbox{\textwidth}{
Gong-Bo Zhao$^{1,2}$\thanks{Email: gbzhao@nao.cas.cn}, Yuting Wang$^{1,2}$, Ashley J. Ross$^{3,2}$, Sarah Shandera$^{4,5}$, Will J. Percival$^{2}$, Kyle S. Dawson$^{6}$, Jean-Paul Kneib$^{7,8}$, Adam D. Myers$^{9}$, Joel R. Brownstein$^{6}$, Johan Comparat$^{10,11}$, Timoth\'ee Delubac${^7}$, Pengyuan Gao$^{1}$,  Alireza Hojjati$^{12,13}$, Kazuya Koyama$^{2}$, Cameron K. McBride$^{14}$, Andr\'es Meza$^{15}$, Jeffrey A. Newman$^{16}$, Nathalie Palanque-Delabrouille$^{17}$, Levon Pogosian$^{13}$, Francisco Prada$^{10,18,19}$, Graziano Rossi$^{20}$, Donald P. Schneider$^{4,21}$, Hee-Jong Seo$^{22}$, Charling Tao$^{23,24}$, Dandan Wang$^{1}$, Christophe Y\`eche$^{17}$, Hanyu Zhang$^1$, Yuecheng Zhang$^{1}$, Xu Zhou$^{1}$, Fangzhou Zhu$^{25}$, Hu Zou$^{1}$}
\vspace*{15pt} \\
$^{1}$ National Astronomy Observatories, Chinese Academy of Science, Beijing, 100012, P.R.China\\
$^{2}$ Institute of Cosmology \& Gravitation, University of Portsmouth, Dennis Sciama Building, Portsmouth, PO1 3FX, UK\\
$^{3}$ Center  for  Cosmology  and  Astro-Particle  Physics,  Ohio State University, Columbus, OH 43210, USA\\
$^{4}$ Institute for Gravitation and the Cosmos, The Pennsylvania State University, University Park, PA 16802, USA\\
$^{5}$ Perimeter Institute for Theoretical Physics, Waterloo, Ontario, Canada \\
$^{6}$ Department of Physics and Astronomy, University of Utah, 115 S 1400 E, Salt Lake City, UT 84112, USA\\
$^{7}$ Laboratoire d'Astrophysique, Ecole polytechnique Fed\'erale de Lausanne, CH-1015 Lausanne, Switzerland\\
$^{8}$ Aix Marseille Universit\'e, CNRS, LAM (Laboratoire d'Astrophysique de Marseille) UMR 7326, F-13388, Marseille, France\\
$^{9}$ Department of Physics and Astronomy, University of Wyoming, Laramie, WY 82071, USA \\
$^{10}$ Instituto de F\'{\i}sica Te\'orica, (UAM/CSIC), Universidad Aut\'onoma de Madrid, Cantoblanco, E-28049 Madrid, Spain\\
$^{11}$ Severo Ochoa IFT Fellow \\
$^{12}$ Department of Physics and Astronomy, University of British Columbia, Vancouver, V6T 1Z1, BC, Canada \\
$^{13}$ Physics Department, Simon Fraser University, Burnaby, V5A 1S6, BC, Canada \\
$^{14}$ Harvard-Smithsonian Center for Astrophysics, Cambridge, MA 02138, USA\\
$^{15}$ Departamento de Ciencias Fisicas, Universidad Andres Bello, Avda. Republica 220, Santiago, Chile\\
$^{16}$ Department of Physics and Astronomy and PITT PACC, University of Pittsburgh, 3941 O'Hara Street, Pittsburgh, PA 15260, USA\\
$^{17}$ CEA, Centre de Saclay, IRFU/SPP, F-91191 Gif-sur-Yvette, France \\
$^{18}$ Campus of International Excellence UAM+CSIC, Cantoblanco, E-28049 Madrid, Spain\\
$^{19}$ Instituto de Astrof\'{\i}sica de Andaluc\'{\i}a (CSIC), Glorieta de la Astronom\'{\i}a, E-18080 Granada, Spain\\
$^{20}$ Department of Astronomy and Space Science, Sejong University, Seoul 143-747, Korea \\
$^{21}$ Department of Astronomy and Astrophysics, The Pennsylvania State University, University Park, PA 16802, USA \\
$^{22}$ Department of Physics and Astronomy, Ohio University, 251B Clippinger Labs, Athens, OH 45701, USA\\
$^{23}$ Centre de Physique des Particules de Marseille, CNRS / IN2P3-Luminy and Universit\'e de la M\'editerran\'ee, \\
~~~~Case 907, F-13288 Marseille Cedex 9, France\\
$^{24}$ Department of Physics and Tsinghua Center for Astrophysics, Tsinghua University, Beijing, 100084, P.R.China\\
$^{25}$ Department of Physics, Yale University, New Haven, CT 06511, USA\\
}
\date{\today} 
\pagerange{\pageref{firstpage}--\pageref{lastpage}}

\label{firstpage}

\maketitle

\clearpage

\begin{abstract}

We present a science forecast for the eBOSS survey. Focusing on discrete tracers, we forecast the expected accuracy of the baryonic acoustic oscillation (BAO), the redshift-space distortion (RSD) measurements, the $f_{\rm NL}$ parameter quantifying the primordial non-Gaussianity, the dark energy and modified gravity parameters. We also use the line-of-sight clustering in the Ly-$\alpha$ forest to constrain the total neutrino mass. We find that eBOSS LRGs, ELGs and Clustering Quasars (CQs) can achieve a precision of 1\%, 2.2\% and 1.6\%, respectively, for spherically averaged BAO distance measurements. Using the same samples, the constraint on $f\sigma_8$ is expected to be 2.5\%, 3.3\% and 2.8\% respectively. For primordial non-Gaussianity, eBOSS alone can reach an accuracy of $\sigma(f_{\rm NL})\sim10-15$. eBOSS can at most improve the dark energy Figure of Merit (FoM) by a factor of $3$ for the Chevallier-Polarski-Linder (CPL) parametrisation, and can well constrain three eigenmodes for the general equation-of-state parameter. eBOSS can also significantly improve constraints on modified gravity parameters by providing the RSD information, which is highly complementary to constraints obtained from weak lensing measurements. A principle component analysis (PCA) shows that eBOSS can measure the eigenmodes of the effective Newton's constant to 2\% precision; this is a factor of 10 improvement over that achievable without eBOSS. Finally, we derive the eBOSS constraint (combined with Planck, DES and BOSS) on the total neutrino mass, $\sigma(\Sigma m_{\nu})=0.03 {\rm eV}$ (68\% CL), which in principle makes it possible to distinguish between the two scenarios of neutrino mass hierarchies.

\end{abstract}


\section{Introduction}
\label{sec:intro}

The cosmic acceleration discovered at the end of last century is one of the most challenging problems to solve in modern science \citep{Riess,Perlmutter}. Possible solutions include introducing dark energy, a hypothetical new energy component in the Universe with a negative pressure (see \citealt{DEreview} for a recent review of dark energy), and modifying general relativity on cosmological scales (see \citealt{MGreview} for a recent review). 

The nature of dark energy and gravity remains unknown, but new observations can provide important information to reveal the underlying fundamental physics. For example, we can infer the nature of dark energy by probing its equation-of-state (EoS) $w(z)$, which is the ratio between its pressure and energy density, and is a function of redshift $z$ in general. In the $\Lambda$CDM model, which is regarded as the standard cosmological model, dark energy is assumed to be the vacuum energy with $w=-1$. Any deviation of $w$ from $-1$, if revealed by observations, might suggest that the dark energy dynamically evolves with time, which will have a significant impact on many subjects in physics. The behaviour of $w$ affects the expansion history of the Universe, thus it can be probed by distance measurements, such as those obtained by measuring the baryonic acoustic oscillations (BAO) signal imprinted on the galaxy clustering pattern on scales of about 150 Mpc. 
Modification of gravity (MG), on the other hand, can give rise to an accelerating Universe without dark energy. In this scenario, MG is predicted to alter the structure formation of the Universe. Thus, if one were to measure a scale-dependent growth pattern on sub-Horizons scales, which is not present in GR, it would be a `smoking gun' for the discovery of MG. Thus the redshift space distortions (RSD) signal \citep{RSD} measured by galaxy surveys is a powerful tool to test gravity.  

Weighing neutrinos is one of the key science drivers of many high-energy experiments. However, due to the tiny cross-section of neutrinos, it is difficult for these experiments to measure the absolute mass of neutrinos. Instead, only the mass differences between neutrino species have so far been measured through neutrino oscillations. Latest measurements give the squared mass differences $\Delta m_{21}^2=7.53\pm0.18\times10^{-5}$\,eV$^2$ and $\Delta m_{32}^2=2.44\pm0.06\times10^{-3}$\,eV$^2$ for the normal mass hierarchy (NH; $m_3\gg m_2 \simeq m_1$) and $\Delta m_{32}^2=2.52\pm0.07\times10^{-3}$\,eV$^2$ for the inverted mass hierarchy (IH; $m_3\ll m_2 \simeq m_1$) \citep{PDG2014}, where $m_1,m_2$ and $m_3$ denote the mass of three different species of neutrinos. Our Universe is an ideal laboratory to measure the total mass of neutrinos and distinguish between two mass hierarchies because massive neutrinos affect cosmological observables in significant ways. Existing in the form of radiation in the early Universe, neutrinos shift the epoch of the matter-radiation equality thus changing the shape of the cosmic microwave background (CMB) angular power spectrum. At late times, massive neutrinos can damp the formation of cosmic structure on small scales due to the free-streaming effect, thus affecting the cosmic growth factor, which can be probed by redshift surveys \citep{Dolgov02,Lesgourgues06}.

Different inflation models predict varying levels of primordial non-Gaussianity (NG), so measuring the NG observationally can test our assumptions of the physical mechanism governing the early Universe. Primordial non-Gaussianity can change the clustering pattern of galaxies on large scales of the Universe through an induced large-scale bias \citep{NGb1,NGb2}. Therefore observing the large scale clustering of galaxies can shed light on the physics in the early Universe.  

The Baryon Oscillation Spectroscopic Survey (BOSS)\footnote{To avoid confusion with the numerous acronyms used in this work, we included a mini-dictionary in Table \ref{tab:acronym}.} \citep{BOSS}, part of the Sloan Digital Sky Survey III (SDSS-III; \citealt{SDSS3}), has observed spectra of more than 1.5 million galaxies brighter than $i=19.9$ and approximately 170,000 new quasars of redshift $2.1\leq3.5$ to a depth of $g<22$ (Paris et al., 2015, in preparation) \footnote{More details of the BOSS filter, spectrograph and pipeline, see \citet{ugrizFilter,BOSSspectro,BOSSpipeline}.}. The precision of BAO and RSD measurements from Data Release 11 (DR11) of BOSS have been reduced to 1-2\% and 6\% respectively, and have provided stringent constraints on dark energy, modified gravity, neutrino mass, primordial non-Gaussianity and other cosmological parameters when combined with other observations \citep{fNL_Ross,mnu_GB,BAODR11,RSDDR11,mnu_FB,mnu2015,mnulya2,mnuGR}. 

The extended Baryon Oscillation Spectroscopic Survey (eBOSS) is a new redshift survey within SDSS-IV, observations for which started in July 2014 \footnote{\url{http://www.sdss.org/surveys/eboss/}}. The eBOSS cosmology program uses the same 1000-fiber optical spectrographs installed on the 2.5 m-aperture Sloan Foundation Telescope \citep{SDSStelescope} at the Apache Point Observatory (APO) in New Mexico, used for the Baryon Oscillation Spectroscopic Survey (BOSS) of SDSS-III. The eBOSS program will map the Universe over the redshift range $0.6<z<2.2$ by observing multiple tracers including luminous red galaxies (LRGs), emission line galaxies (ELGs) and quasars: a sample that combines eBOSS LRGs with the BOSS LRGs at $z>0.6$ provides a 1\% distance measurement; the ELGs sample offers a 2\% estimate at slightly higher redshifts; and the clustering quasars (CQs) produce a 1.6\% measurement in $0.9<z<2.2$ \citep{Overview} \footnote{The clustering quasars provide a 2\% BAO distance measurement if 58 quasars per square degree over $0.9<z<2.2$ is assumed.}. These distance measurements are expected to improve the dark energy Figure of Merit (FoM; \citealt{DETF}) by a factor of $3$ compared to BOSS results. 

This paper presents the expected cosmological implications of the eBOSS survey including the BAO and RSD measurements and $f_{\rm NL}$ constraints, and is one of a series of technical papers describing the eBOSS survey. In Section 2, we describe the eBOSS survey in details. We outline the methodology used for the science forecasts for discrete tracers in Section 3. Our forecasts on cosmological parameters also include the expected BAO-scale precision from the 3D Lyman Alpha Forest (Ly-$\alpha$) clustering. We present the results in Section 4. Section 5 contains conclusions and discussions. 

\begin{table}
\begin{center}
\begin{tabular}{|c|c|}
\hline \hline 
The acronym & The meaning \\ 
\hline 
APO & Apache Point Observatory \\
BAO & Baryon Acoustic Oscillation \\ 
BOSS &  Baryon Oscillation Spectroscopic Survey \\
CMB & Cosmic Microwave Background \\ 
CPL & Chevallier-Polarski-Linder \\
CQs & Clustering Quasars \\
DECam & Dark Energy Survey camera \\
FoM & Figure of Merit \\ 
DE & Dark Energy \\ 
DES & Dark Energy Survey \\ 
DESI & Dark Energy Spectroscopic Instrument \\
DR & Data Release \\ 
eBOSS &   extended Baryon Oscillation Spectroscopic Survey \\
ELGs & Emission Line Galaxies \\
EoS & Equation-of-State \\
FoG & Fingers-of-God \\ 
GR & General Relativity \\  
LRGs & Luminous Red Galaxies \\   
MG & Modified Gravity \\ 
NG & non-Gaussianity \\ 
RSD & Redshift Space Distorsion \\ 
SCUSS & South Galactic Cap U-band Sky Survey \\
SDSS & Sloan Digital Sky Survey \\
WISE &  Wide-Field Infrared Survey Explorer\\
WL & Weak Lensing \\ 
\hline \hline 
\end{tabular}
\end{center}
\label{tab:acronym}
\caption{The acronyms used in this work and their expression.}
\end{table}%

\section{The eBOSS survey}
\label{sec:survey}

The eBOSS survey is described in detail in \citet{Overview}, and we highlight the key facts here. 

Motivated by the success of BOSS, eBOSS will extend the SDSS BAO measurement to ${0.6<z<1}$ using LRGs and ELGs, and make the first BAO measurement at ${0.9<z<2.2}$ using quasars.  

The selected LRGs will cover the redshift range of ${0.6<z<1}$ over 7000 deg$^2$ with a surface number density of 50 deg$^{-2}$. We assume a bias model of $ b(z)_{\rm LRG}=1.7G(0)/G(z)$, where $G(z)$ is the linear growth factor at redshift $z$. Details of LRGs target selection are presented in \citet{LRGtech}.

The ELGs survey will start in Fall of 2016. The target selection definitions of the ELGs sample are not yet finalised and thus we explore three possible selection options, each of which will use some subset of the following imaging data: the South Galactic Cap U-band Sky Survey (SCUSS) \citep{scuss1,scuss2}\footnote{For more information about the SCUSS survey, see \url{http://batc.bao.ac.cn/Uband/}}, Sloan Digital Sky Survey (SDSS) $griz$ \citep{ugrizFilter}, Wide-Field Infrared Survey Explorer (WISE) \citep{WISE}, or $grz$ imaging with the Dark Energy Survey camera (DECam) \citep{DECAM} \footnote{Another option is mentioned in the eBOSS overview paper \citep{Overview}, which only uses the $gri$ and $Uri$ bands of the SDSS and SCUSS imaging for target selection. We are not including it here because it produces tracers at low efficiency (only 52.5\%).}. The proposed selections are:
\begin{itemize}
\item {Fisher Discriminant}: The targets are selected using the WISE, SDSS and SCUSS photometry with a cut on the Fisher discriminant quantities instead of cuts in the colour-colour diagrams \citep{ELG_Fisher}. The initial tests of this scheme demonstrate its validity: it approaches the requirement that 74\% of targets turn out to be ELGs in the redshift range $0.6<z<1.0$ (henceforth referred to as the ``74\% putiry requirement''). We assume a completeness of 95\% over 1500 deg$^2$;
\item {Low Density DECam}: The targets are selected from DECam $grz$ photometry. The deeper photometry means this selection exceeds the 74\% purity requirement for $0.7<z<1.1$. The expected target density is $\sim190$ deg$^{-2}$. The survey area is assumed to be 1400 deg$^{-2}$;
\item {High Density DECam}: The targets are selected in a similar way to the `Low Density' case but the colour cuts are tuned to achieve a target density of $\sim240$ deg$^{-2}$ over 1100 deg$^{-2}$ \citep{Overview}.
\end{itemize}

We assume a bias of $b(z)_{\rm ELG}=1.0G(0)/G(z)$ for the ELGs \citep{Overview}.

The clustering quasars will be targeted using the XDQSOz algorithm \citep{XDQSO}, which was used for the quasar sample of BOSS, applied on the {\tt QSO\_CORE} sample in eBOSS. The expected number density to obtain 2\% prevision on the BAO measurement over the redshift range $0.9<z<2.2$ is 58 deg$^{-2}$ over an area of 7500 deg$^2$. This number is quoted as the base requirement for the CQs in \citet{Overview} and \citet{QSOtech}. In reality, the eBOSS selection approach detailed in \citet{QSOtech} exceeds this metric, successfully targeting closer to 70 deg$^2$ $0.9 < z < 2.2$ quasars over 7500 deg$^2$. In the rest of this paper (\ie, see Table 1), we adopt the redshift distribution corresponding to this expected quasar density of 70 deg$^2$ 0.9 < z < 2.2 from \citet{QSOtech}. This selection contains a useful tail of an additional $\sim$ 8\,deg$^{-2}$ quasars in the redshift range $0.6 < z < 0.9$, which we include in our forecasts throughout the rest of this paper. Note that in \citet{QSOtech} the CQs are referred to as the {\tt QSO\_CORE} sample.

\begin{figure}
\centering
{\includegraphics[scale=0.18]{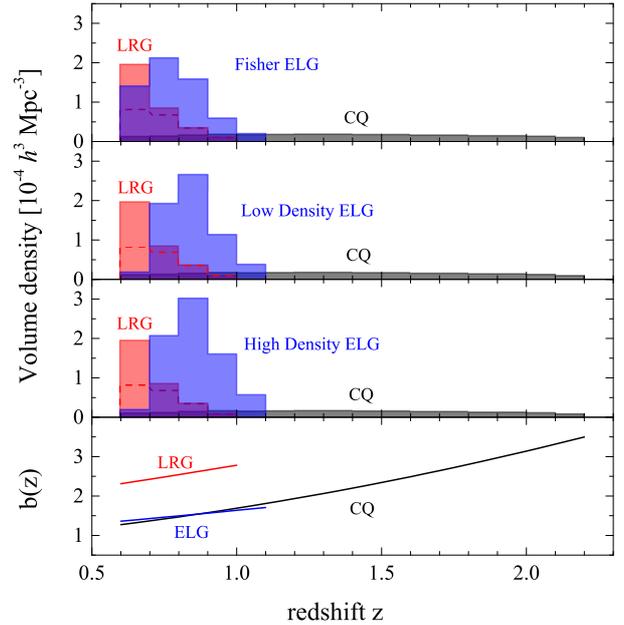}}
\caption{The volume number density (in units of $10^{-4}\vunits$) and galaxy bias of the LRGs (red), ELGs (blue) and clustering quasars (black). In the three upper panels, the red dashed lines show the eBOSS LRGs, and the red shaded region show the eBOSS LRGs combined with the BOSS LRGs in the $z>0.6$ tail.}
\label{fig:nz}
\end{figure}

We assume the bias of the clustering quasars to be $b(z)_{\rm CQ}= 0.53+0.29(1+z)^2$ \citep{Croom05,QSObias}.
Table \ref{tab:nz} summarises the targets used in this work, including the number and volume number density of each type of targets in each redshift slice, the effective redshift, the total number of targets, the surface area and the bias. We follow \citet{Overview} and take a conservative sky area for the LRGs to be 7000 deg$^2$ instead of 7500 deg$^2$. Different tracers overlap maximally in the survey area.
Fig \ref{fig:nz} shows the redshift distribution we adopt for the tracers, where the overlap in redshifts is apparent. The time evolution of the biases is also shown.  

\begin{table*}

\centering
\begin{tabular}{l c c c  c c c}
\hline\hline
Redshift  & CMASS & eBOSS & Clustering & Fisher & Low Density & High Density \\ 
& LRGs & LRGs   & Quasars & ELGs & DECam ELGs & DECam ELGs   \\ 

\hline
$0.6 < z < 0.7$ & { 137,475 (1.137)} & { 97,937 (0.810)} &  {15,416 (0.119)}   & { 36,584 (1.412)} & {4,425 (0.183)} & {3,895 (0.205)}   \\
$0.7 < z < 0.8$ & { 24,407 (0.170)} & { 97,340 (0.678)}   &  {19,997 (0.130)}   & { 66,606 (2.165)} & { 54,786 (1.908)} & {46,656 (2.068)}   \\
$0.8 < z < 0.9$ & { 1,645 (0.010)} & { 57,600 (0.350)}    & {27,154 (0.154)}   & { 58,328 (1.654)} & { 87,979 (2.673)} & {78,462 (3.034)}   \\
$0.9 < z < 1.0$ & { 183 (0.001)} & { 17,815 (0.097)}       & { 33,649 (0.171)}   & { 24,557 (0.624)} & { 41,690 (1.135)} & {46,321 (1.605)}   \\
$1.0 < z < 1.1$ &  			&  			            & { 35,056 (0.163)}             & {9,377 (0.218)} & { 14,975 (0.373)} & {17,917 (0.568)}   \\
$1.1 < z < 1.2$ &  			&  				    & { 39,307 (0.170)}             & {3,736 (0.081)} &{ 6,863 (0.159) }& { 8,173 (0.241)}   \\
$1.2 < z < 1.4$ &  			&  			            & {87,984 (0.175)}  &  &  &    \\
$1.4 < z < 1.6$ &  			&			 	    & { 90,373 (0.166)}  &  &  &    \\
$1.6 < z < 1.8$ &  			&  				   & { 86,631 (0.151)}  &  &  &    \\
$1.8 < z < 2.0$ &  			&  				  & { 81,255 (0.137)}  &  &  &    \\
$2.0 < z < 2.1$ &  			&  			  	& { 36,760 (0.122)}  &  &  &    \\
$2.1 < z < 2.2$ & 			 & 			  	& { 28,214 (0.093)} &  & &    \\
\hline
$z_{\rm eff}$& 0.665 & 0.736 & 1.374 & 0.790 & 0.851& 0.863 \\
\hline
Total & {163,710 (0.267)} & {270,692 (0.442)} & {581,796 (0.148)}& {199,188(0.903)} & {210,718 (1.024)} & {201,424 (1.245)} \\
\hline
Surface Area & 7000 $\sqdeg$ & 7000 $\sqdeg$  & 7500 $\sqdeg$ & 1500 $\sqdeg$ & 1400 $\sqdeg$ & 1100 $\sqdeg$ \\
\hline
Bias                                &$1.7\frac{G(0)}{G(z)}$ & $1.7\frac{G(0)}{G(z)}$ & $0.53+0.29(1+z)^2$ &$\frac{G(0)}{G(z)}$&$\frac{G(0)}{G(z)}$  & $\frac{G(0)}{G(z)}$ \\

\hline\hline
\end{tabular}\caption{
Expected number of each target class in each redshift bin, and the volume density in units $10^{-4}\vunits$ shown in parentheses. The effective redshift $z_{\rm eff}$, total number of sources, observed surface area and galaxy bias of each target is shown in the last four rows.}
\label{tab:nz}
\end{table*}

\section{Methodology}
\label{sec:method}

In this section, we present the methodology for the science forecast, which is based on the Fisher matrix technique \citep{Fisher}. The formalism is presented in sections \ref{sec:FisherPk} - \ref{sec:FCMBWL}, and the parametrisation and fiducial cosmology is shown in Sec. \ref{sec:para}.  We allow for the multi-tracer nature of eBOSS, including the cross-correlation when using the power spectra of different kinds of targets in overlapping regions of sky and in redshift.  

\subsection{The Fisher matrix for $P(k)$ of redshift surveys}
\label{sec:FisherPk}

Using the 3D galaxy power spectrum in redshift space measured from eBOSS, the Fisher matrix element for a pair of arbitrary parameters $\{p_i,p_j\}$ is given by \citep{FisherPk} \footnote{Note that this is the Fisher matrix using galaxies distributed in a given redshift slice. The final Fisher matrix is the sum over the Fisher matrices of individual redshift bins.}, \ba \label{eq:Fisher} &&F_{ij}=\frac{V_{\rm sur}}{4{\pi}^2}\int_{-1}^{1} {\rm d}\mu \int_{k_{\rm min}}^{k_{\rm max}}  k^2 {\rm d}k~\mathcal{F}_{ij}(k,\mu), \\ 
 &&  k_{\rm min}=\frac{2\pi}{V_{\rm sur}^{1/3}} \ [{\rm h/Mpc}], \ \ k_{\rm max}=0.1 \frac{D(0)}{D(z)} \ \ [{\rm h/Mpc}], \\
\label{eq:F_matrix} && \mathcal{F}_{ij}(k,\mu)=\frac{1}{2}\mathrm{Tr}[{\bf C}_{,i}{\bf C}^{-1}{\bf C}_{,j}{\bf C}^{-1}]. \ea
where $V_{\rm sur}$ is the volume of the redshift survey, $k$ denotes the amplitude of mode ${\bf k}$, $\mu$ is the cosine of the angle between mode ${\bf k}$ and the line of sight, and $D(z)$ is the growth function at redshift $z$. ${\bf C}$ is the data matrix storing the observed galaxy power spectra $P$ in redshift space \footnote {We include the damping term in the power spectra to account for the Fingers-of-God (FoG) effect and for the redshift errors.}, and ${\bf C}_{,i}$ is the derivative matrix with respect to parameter $p_i$, 
 As eBOSS target classes overlap, we shall explicitly show the Fisher matrix for the single- and multi-tracer cases in what follows.

\subsubsection{The single-tracer case}
\label{sec:1tracer}

If there is only one tracer being surveyed, which is the case for most of the sky covered by eBOSS,
\be\label{eq:C_single} C={P}+\frac{1}{n}, \ P = (b+f\mu^2)^2 P_m(k), \ee where $n$, $b$, $f$, $P_m$ denotes the number density, bias, logarithmic growth rate and the matter power spectrum, respectively. In this case, Eq (\ref{eq:F_matrix}) reduces to,
\be\label{eq:F_single} \mathcal{F}_{ij}(k,\mu)=\frac{1}{2}D_{i}D_{j} R_{V}.\ee
\be D_{i}=\frac{\partial \ {\rm ln}{P}}{\partial p_i}, \ R_V\equiv\frac{V_{\rm eff}}{V_{\rm sur}}=\left(\frac{n{P}}{n{P}+1} \right)^2  \ee where ${P}, V_{\rm eff}$ denote the power spectrum in redshift space and the effective volume respectively.

\subsubsection{The double-tracer case}
\label{sec:2tracer}

If two tracers with different biases (denoted by A and B) are used to probe the same patch of the sky in the same redshift range, \eg, the eBOSS LRGs and ELGs, we need to include the cross-correlation, denoted by $X$, between them. In this case, 
${\bf C}$ becomes a $2\times2$ matrix, namely,

\ba
{\bf C}=\begin{bmatrix}
    {P}_{{\rm A}}+\frac{1}{n_{\rm A}} & {P}_{\rm X} \\
    {P}_{\rm X}&{P}_{{\rm B}}+\frac{1}{n_{\rm B}}\\
  \end{bmatrix}
\ea  

The Fisher matrix can be calculated by substituting ${\bf C}$ into Eq (\ref{eq:F_matrix}), and we include an explicit calculation for the 2-tracer case in the Appendix. 
 
\begin{table*}
\begin{center}
\begin{tabular}{cccc}

\hline\hline
& Parameter                                          &  Meaning                                              &   Fiducial value \\ \hline
&${\rm ln}(D_A/s)(z_i)$                                &  The transverse BAO distance for the $i$th redshift bin  & The value for the fiducial cosmology  \\
{${\rm\bf P_I}$}&${\rm ln}(sH)(z_i)$                            &  The line-of-sight BAO distance for the $i$th redshift bin      & The value for the fiducial cosmology  \\
\hline
&$f(z_i)\sigma_8(z_i)$                          & The product of the logarithmic growth and $\sigma_8$  for the $i$th redshift bin     &  The value for the fiducial cosmology  \\
{ ${\rm\bf P_{II}}$}&$b_A(z_i)\sigma_8(z_i)$                     & The product of the bias factor and $\sigma_8$  for the $i$th redshift bin                             &  The value for the fiducial cosmology  \\
 \hline 
&$\omega_{b}\equiv\Omega_{b}h^2$  & The physical baryon energy density & $0.022242$        \\
&$\omega_{c}\equiv\Omega_{c}h^2$  & The physical dark matter energy density & $0.11805$        \\
&$H_0$  & The Hubble constant [km/s/Mpc] & $68.14$       \\
{ ${\rm\bf P_{III}}$}&$\tau$  & The optical depth & $0.0949$        \\
&log$[10^{10}A_s]$  & The amplitude of the primordial power spectrum & $3.098$        \\
&$n_s$  & The spectral index of the primordial power spectrum & $0.9675$        \\
\hline
&$\Sigma m_{\nu}$  & The sum of the neutrino masses in the unit of eV & $0.06$        \\
&$w_0$  & The $w_0$ parameter in the CPL parametrisation          & $-1$\\
&$w_a$  & The $w_a$ parameter in the CPL parametrisation         & $0$\\
{ ${\rm\bf P_{IV}}$}&$w_i$  & The equation-of-state parameter of dark energy in the $i$th redshift bin & $-1$ \\
&$\mu_{ij}$  & The effective Newton's constant in the $\{i,j\}$th pixel in the $\{k,z\}$ plane & $1$ \\
&$\eta_{ij}$  & The gravitational slip in the $\{i,j\}$th pixel in the $\{k,z\}$ plane & $1$ \\
&$f_{\rm NL}$  & The non-Gaussianity parameter & $0$ \\
&$\alpha$ & The power index for the general non-Gaussianity model & $2$ \\

\hline\hline

\end{tabular}
\end{center}
\caption{The parameters used in our forecast, their physical meaning, and the fiducial values we choose, which are consistent with the Planck cosmology \citep{Planck2015}.}
\label{tab:param}
\end{table*}%

\begin{figure}
\centering
{\includegraphics[scale=0.27]{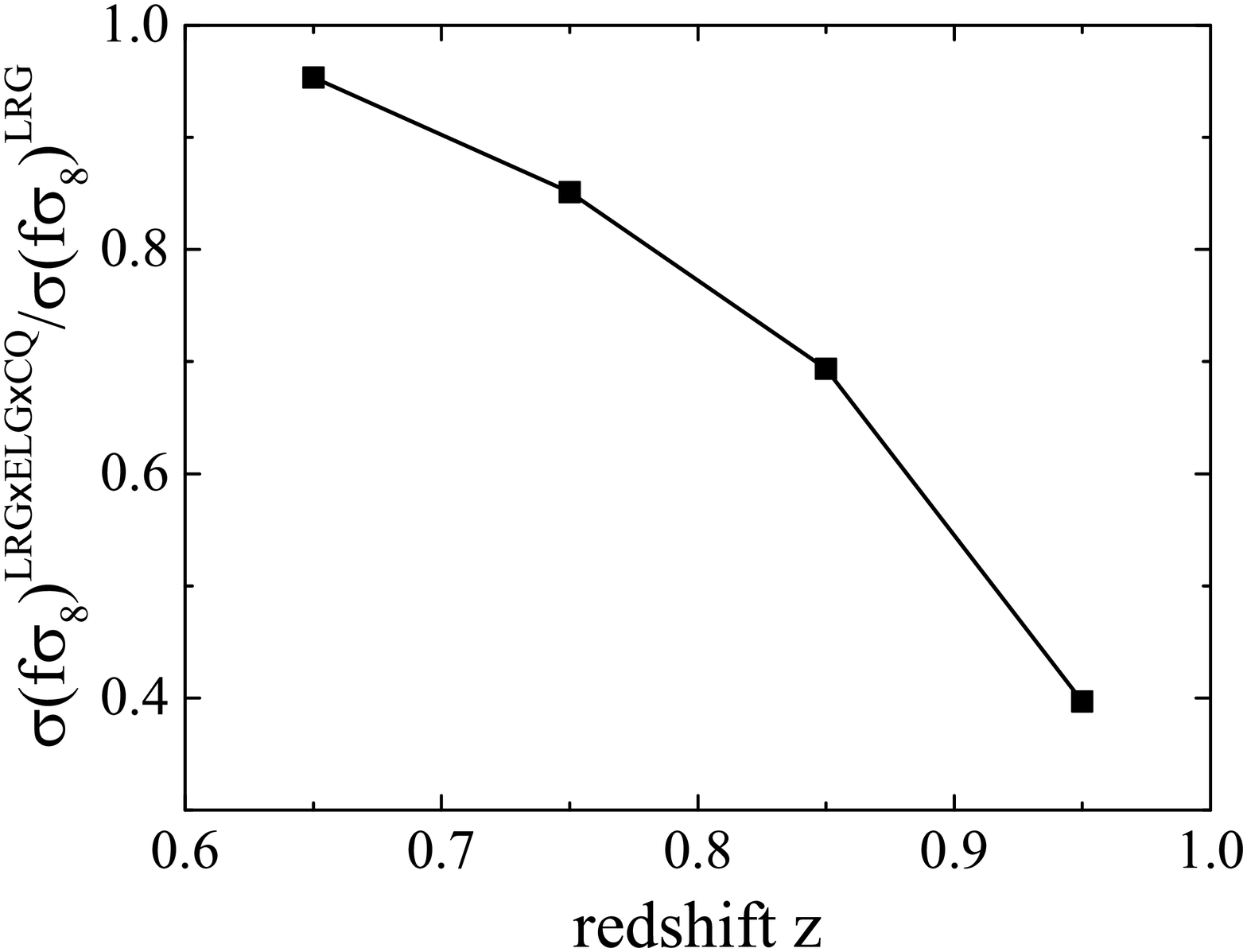}}
\caption{The ratio of the error of $f\sigma_8$ using all the eBOSS tracers to that using eBOSS LRGs alone.}
\label{fig:fs8_2to1}
\end{figure}

\begin{figure}
\centering
{\includegraphics[scale=0.27]{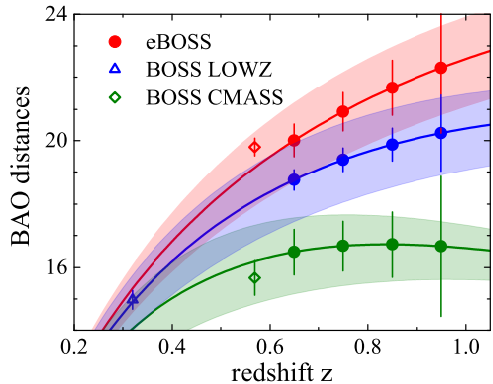}}
\caption{The BAO distance measurement using the eBOSS LRGs, in comparison with that using the BOSS LOWZ and CMASS samples. The top red, middle blue and bottom green data and error bands show ${(1+z)D_A(z)/r_d \sqrt{z}}$, ${D_V(z)/(r_d \sqrt{z})}$ and ${c \sqrt{z} /[H(z) r_d]}$ respectively. The $\sqrt{z}$ factor is included to tune the dynamical ranges for the purpose of visualisation. }
\label{fig:BAOLRG}
\end{figure} 


Compared to the single-tracer case, the auto- and cross-power spectra of multiple tracers provide measurements of ratios of $f/b$ that couple radial and angular modes, beating radial sample variance in the low-noise limit \citep{multi-tracer2}. 

To illustrate the improvement of having multiple tracers, we show an example of the $f\sigma_8$ constraint using eBOSS LRGs and all the eBOSS tracers. We start by constraining $f\sigma_8$ using the eBOSS LRGs, and will show the improvement of the constraint when the ELGs and CQs are added to the analysis, with the cross-correlation included. The result is shown in Fig \ref{fig:fs8_2to1}, in which the ratio of the error on $f\sigma_8$ using multiple tracers to that using the eBOSS LRGs alone is plotted as a function of redshift. 

As shown, the $f\sigma_8$ constraint will be improved when a full cross-correlation analysis among the ELGs, CQs and LRGs is possible in 2020. The improvement is maximal in the redshift bin of $0.9<z<1.0$, reducing the $f\sigma_8$ uncertainty from $13.7\%$ to $5.4\%$. 

Note that this improvement is mainly due to the fact that more galaxies are available in the full cross-correlation analysis. Although the gain from the reduction of sample variance is sub-dominant for the case of eBOSS due to the level of shot noise, we show there will be a clear benefit from combining all of the available samples and their cross-correlations. Further, we expect that using multiple tracers in the overlapping volume will be helpful to diagnose and reduce both observational and theoretical systematic uncertainties.  

\begin{table*}
\begin{center} 
\begin{tabular}{cccccccc}
\hline\hline
    Sample   & redshift    & $\bar{n}P_{0.2,0}$ & $\sigma_{D_A}/D_A $  & $\sigma_{H}/H $  &$\sigma_{D_V}/D_V$  & $\sigma_{f\sigma_8}/f\sigma_8 $ & $\sigma_{b\sigma_8}/b\sigma_8$   \\ \hline
                   &$0.6<z<0.7$     &  0.684& 0.030 & 0.049 & 0.020 & 0.048  &  0.007          \\  
                   &$0.7<z<0.8$    & 0.102 & 0.073 & 0.106 & 0.047 & 0.104 &  0.016          \\  
CMASS LRGs
                   &$0.8<z<0.9$    & 0.006 & 0.830 & 1.109 & 0.523 & 1.083 &  0.173          \\  
                   &$0.9<z<1.0$   & 0.0006  &7.439&  9.955  &4.690& 9.936 & 1.557           \\
                   & ${\bf 0.6<z<1.0}$ ${\bf (z_{\rm eff}=0.665)}$   & {\bf 0.161}  &{\bf 0.027}&  {\bf 0.040}  &{\bf 0.017}& {\bf 0.039} & {\bf 0.006  }         \\  \hline    
                   &$0.6<z<0.7$    &0.487  & 0.034 & 0.054 & 0.022 & 0.053 & 0.008           \\  
                   &$0.7<z<0.8$    &0.408  & 0.032 & 0.050 & 0.021 & 0.051 & 0.007           \\  
eBOSS LRGs                    &$0.8<z<0.9$  & 0.211 & 0.041 & 0.062 & 0.026 & 0.063 &  0.009          \\  
                   &$0.9<z<1.0$   & 0.058 & 0.094 & 0.134 & 0.060 & 0.137 & 0.021           \\    
                   &${\bf 0.6<z<1.0}$ ${\bf (z_{\rm eff}=0.736)}$   &  {\bf 0.266 } & {\bf0.019}&   {\bf0.030}  & {\bf0.013}&  {\bf0.029} &  {\bf0.004 }          \\  \hline    
                   &$0.6<z<0.7$     & 1.172 &0.026  & 0.043 & 0.017 &0.043  &  0.006          \\  
                   &$0.7<z<0.8$    & 0.510 & 0.029 & 0.046 & 0.019 & 0.047 &  0.007          \\  
CMASS+eBOSS LRGs                    &$0.8<z<0.9 $   & 0.217 & 0.040 & 0.061 & 0.026 & 0.062 &   0.009         \\  
                   &$0.9<z<1.0 $   & 0.059 & 0.093 & 0.133 & 0.059 & 0.136 &   0.020         \\       
                   &${\bf 0.6<z<1.0}$ ${\bf (z_{\rm eff}=0.707)}$   & {\bf 0.427}  &{\bf 0.016} & {\bf 0.025}  & {\bf 0.010}& {\bf 0.025} &{\bf  0.004 }          \\  \hline    
                   &$0.6<z<0.7 $   & 0.294   & 0.090 & 0.130  & 0.058 & 0.085 &  0.021      \\  
                   &$0.7<z<0.8 $   & 0.451  & 0.065 &  0.097 &  0.042 & 0.065  &   0.015      \\  
                   &$0.8<z<0.9 $   & 0.344  & 0.064 &  0.094 &  0.041 & 0.064 &   0.015         \\  
Fisher ELGs                     &$0.9<z<1.0 $   & 0.130 & 0.105 &  0.140 & 0.066 &  0.098 &  0.024           \\   
                   &$1.0<z<1.1$   & 0.045 & 0.222 &  0.275 &  0.137 &  0.196 &  0.050          \\  
                   &$1.1<z<1.2$   & 0.017 & 0.514& 0.611  &  0.316 &  0.444 & 0.115          \\      
                   &${\bf 0.6<z<1.2}$ ${\bf (z_{\rm eff}=0.790)}$   &{\bf 0.188}  &{\bf 0.037}& {\bf 0.051}  &{\bf 0.023}&{\bf 0.034} &{\bf 0.009}           \\  \hline                                   
                  &$0.6<z<0.7$ &0.381  & 0.038 & 0.458 & 0.238 & 0.299 & 0.084           \\   
                   &$0.7<z<0.8$      &0.397  &0.071  & 0.105 & 0.045 & 0.070 & 0.017         \\  
                   &$0.8<z<0.9 $   & 0.557 & 0.054 & 0.082 & 0.035 & 0.056 & 0.013         \\  
Low Density ELGs                     &$0.9<z<1.0 $    & 0.236 & 0.074 & 0.104 & 0.047 & 0.073 & 0.018             \\  
                   &$1.0<z<1.1 $    &0.078 & 0.149 & 0.191 & 0.093 & 0.137 & 0.034          \\   
                   &$1.1<z<1.2$   &0.033 & 0.286 &0.351  & 0.177 &0.256 &  0.065        \\    
                   &${\bf 0.6<z<1.2}$ ${\bf (z_{\rm eff}=0.851)}$   &{\bf 0.213}  &{\bf 0.035}& {\bf 0.048}  & {\bf 0.022}& {\bf 0.033} & {\bf 0.008 }          \\  \hline    
                   &$0.6<z<0.7$ &0.043  & 0.397 &0.473  & 0.244 & 0.309 & 0.087           \\    
                   &$0.7<z<0.8$     &0.431  & 0.077 & 0.115 & 0.050 & 0.077  & 0.018           \\  
                   &$0.8<z<0.9$   & 0.632 & 0.058 & 0.090 & 0.038 & 0.061 &  0.014          \\  
High Density ELGs                     &$0.9<z<1.0$    & 0.334 & 0.069 & 0.101 & 0.044 & 0.071 & 0.017           \\  
                   &$1.0<z<1.1$  & 0.118 & 0.122 & 0.162 & 0.077 & 0.117 & 0.029           \\  
                    &$1.1<z<1.2$   &0.050 & 0.225 & 0.282 & 0.140 &0.207 & 0.052         \\    
                    &${\bf 0.6<z<1.2}$ ${\bf (z_{\rm eff}=0.863)}$   &{\bf 0.259}  &{\bf0.035}&{\bf  0.050 } &{\bf0.022}& {\bf0.034} &  {\bf0.008  }        \\  \hline      
                   &$0.6<z<0.7$ &0.022  &0.267  & 0.300 & 0.163 &0.189  & 0.058           \\    
                   &$0.7<z<0.8$     & 0.025 & 0.211  & 0.243 & 0.129  & 0.158  & 0.046        \\  
                   &$0.8<z<0.9$    & 0.032 &  0.158 & 0.187  & 0.097 & 0.126 & 0.035          \\ 
                   &$0.9<z<1.0$     &0.037  & 0.127 & 0.155 & 0.079 & 0.109  & 0.028           \\  
                   &$1.0<z<1.1$   & 0.037 & 0.120 & 0.148 & 0.074 & 0.109 &  0.027          \\  
                   &$1.1<z<1.2$   & 0.041 & 0.104 & 0.132 & 0.065 & 0.101 &  0.024          \\  
 Clustering Quasars                     &$1.2<z<1.4$    & 0.045 & 0.063 & 0.082 & 0.039 & 0.067 &  0.014          \\  
                   &$1.4<z<1.6$    & 0.047 & 0.057 & 0.076 & 0.036 & 0.068 &   0.013         \\  
                   &$1.6<z<1.8$   & 0.047 & 0.054 & 0.075 & 0.034 & 0.072 &   0.012         \\  
                   &$1.8<z<2.0$   & 0.047 & 0.052 & 0.074 & 0.033 & 0.078 &   0.012         \\  
                   &$2.0<z<2.1$   & 0.045 &  0.076&  0.108& 0.049 & 0.121 &  0.017          \\
                   &$2.1<z<2.2$   & 0.036 &  0.092&  0.132& 0.059 & 0.153 &  0.021          \\   
                   &${\bf 0.6<z<2.2}$ ${\bf (z_{\rm eff}=1.374)}$   &{\bf 0.040}  &{\bf 0.025}& {\bf 0.033}  &{\bf 0.016}&{\bf 0.028} & {\bf0.006 }           \\  \hline  \hline                       
\end{tabular}
\end{center}
\caption{The predicted 68\% CL error of the BAO distances and RSD parameters using various tracers in different redshift slices. In the last row for each tracer, we show the forecast using the total of all targets distributed across all redshift slices. We also show the effective redshift in parentheses.}
\label{tab:BAORSD}
\end{table*}

\begin{table}

\begin{center}
\begin{tabular}{ccc}
\hline\hline
      Sample   &redshift range &  $\sigma_{f\sigma_8}/f\sigma_8$   \\ \hline
                    &$0.6<z<0.7$      &  0.039              \\   
                    &$0.7<z<0.8$    &    0.039                \\ 
                    &$0.8<z<0.9$     &    0.043               \\
                     &$0.9<z<1.0$     &     0.061           \\  
                     &$1.0<z<1.1$     &     0.088            \\ 
                     &$1.1<z<1.2$     &     0.094             \\   
 Combined I                    &$1.2<z<1.4$     & 0.067          \\  
                   &$1.4<z<1.6$     & 0.068             \\  
                   &$1.6<z<1.8$    & 0.072          \\  
                   &$1.8<z<2.0$    & 0.078          \\  
                   &$2.0<z<2.1$    & 0.121           \\
                   &$2.1<z<2.2$    & 0.153           \\     \hline    
		 &$0.6<z<0.7$      &  0.041              \\   
                     &$0.7<z<0.8$      &   0.040               \\   
                     &$0.8<z<0.9$    &    0.041               \\ 
                      &$0.9<z<1.0$     &   0.054               \\
                      &$1.0<z<1.1$    &     0.080               \\  
                     &$1.1<z<1.2$     &     0.088             \\   
Combined  II                     &$1.2<z<1.4$     & 0.067          \\  
                   &$1.4<z<1.6$     & 0.068             \\  
                   &$1.6<z<1.8$    & 0.072          \\  
                   &$1.8<z<2.0$    & 0.078          \\  
                   &$2.0<z<2.1$    & 0.121           \\
                   &$2.1<z<2.2$    & 0.153           \\     \hline                        
		&$0.6<z<0.7$      &  0.041               \\   
                     &$0.7<z<0.8$    &     0.040                \\   
                     &$0.8<z<0.9$    &     0.043               \\ 
                      &$0.9<z<1.0$     &     0.054               \\
                      &$1.0<z<1.1$     &     0.076                \\
                     &$1.1<z<1.2$     &     0.086            \\   
Combined  III                    &$1.2<z<1.4$     & 0.067          \\  
                   &$1.4<z<1.6$     & 0.068             \\  
                   &$1.6<z<1.8$    & 0.072          \\  
                   &$1.8<z<2.0$    & 0.078          \\  
                   &$2.0<z<2.1$    & 0.121           \\
                   &$2.1<z<2.2$    & 0.153           \\                            
 \hline\hline                                                    
\end{tabular}
\end{center}
\caption{Predictions for the precision of $f\sigma_8$ measurements obtained using the multi-tracer technique, using three different eBOSS data combinations: I. LRGs+Fisher ELG+clustering quasar; II. LRGs+Low density ELG+clustering quasar; III. LRGs+High density ELG+clustering quasar.}
\label{tab:RSDcombo}
\end{table} 

\subsection{The Fisher matrix for the BAO of redshift surveys}
\label{sec:FisherBAO}
To forecast the sensitivity of the BAO distance along and perpendicular to the line of sight for eBOSS, we follow \citet{FisherBAO}.

The two BAO parameters are,
\be
\ln (D_A/s), \ \ln(s H)
\ee

Note that \be \sigma(\ln (D_A/s)) \simeq \ln D_A, \ \sigma(\ln(s H)) \simeq \ln H \ee if the sound horizon $s$ can be determined by external data such as the CMB, which can be achieved for eBOSS using Planck measurements (\eg \ \citealt{Planck2015}).

\subsection{The Fisher matrix for the RSD of redshift surveys}
\label{sec:FisherRSD}

We follow \citet{FisherRSD} to perform forecasts for the RSD parameters. The observable used is the full galaxy power spectrum in redshift space. To be consistent with the notation of \citet{FisherRSD}, we rewrite Eq (\ref{eq:C_single}) as, \be
P=  \left[b \sigma_8(z)+f \sigma_8(z)\mu^2\right]^2 \times\frac{P_m(k,z=0)}{\sigma^2_8(z=0)}
\ee \ie, for each redshift slice, we attach $\sigma_8(z)$ to $b(z)$ and $f(z)$ and use the products as parameters. Explicitly, the free parameters are, \be
\ln [b \sigma_8(z)], \ \ln [f \sigma_8(z)]
\ee
The derivatives of $P$ with respect to these parameters are \footnote{We drop the dependence on $z$ for brevity.}, 
\ba
&&\frac{\partial \ln P}{\partial  \ln (b \sigma_8)}=\frac{2  \ b \sigma_8}{b \sigma_8+f \sigma_8 \ \mu^2}, \\
&&\frac{\partial \ln P}{\partial  \ln (f \sigma_8)}=\frac{2\mu^2 \ f \sigma_8}{b \sigma_8+f \sigma_8 \ \mu^2}.
\ea Note that, in the $N$-tracer case, $p_1$ needs to be extended into a set, namely, \be p_1=\left\{ \ln (b_1 \sigma_8), \ln (b_2 \sigma_8), ... ,\ln (b_N \sigma_8) \right\} \ee

\subsection{The Fisher matrix for the primordial non-Gaussianity}
\label{sec:fNL}

In the context of the local ansatz for non-Gaussianity, where the Bardeen potential $\Phi$ contains a term that is quadratic in a Gaussian field $\phi$, \ie, $\Phi=\phi+f_{\rm NL}(\phi^2-\langle\phi^2\rangle)$, a scale-dependent non-Gaussian bias $\Delta b(k)$ is induced \citep{NGb1,NGb2},
\be
\Delta b(k)=3f_{\rm NL}(b-p)\delta_c\frac{\Omega_m}{k^2T(k)D(z)}\left(\frac{H_0}{c}\right)^2\,, 
\ee
where $p$ depends on the type of tracer \citep{fNL1}, $\delta_c$ is the critical linear over-density for the collapse, and $T(k)$ is the matter transfer function (normalised to unity on large scales). 

The non-Gaussian bias is sensitive to any coupling between modes of very different scales, which could come from bispectra or higher order correlations in models other than the local ansatz. In that sense, the halo bias is an important probe of non-Gaussianity beyond the local ansatz. To constrain non-Gaussian models more generally, the non-Gaussian bias can be parametrised by, 
\ba
\label{eq:generalNGbias}
\Delta b(k) = 3\,\mathcal{A}_{\rm NL}(b-p)\delta_c\frac{\Omega_m}{k^2(k/k_p)^{\alpha-2}T(k)D(z)}\left(\frac{H_0}{c}\right)^2\;.
\ea
By allowing values of $\alpha$ different from 2, this form tests the scaling of the squeezed limit of the bispectrum. The coefficient $\mathcal {A}_{\rm NL}$ may depend on the mass of the object (through the Gaussian bias) depending on the details of the bispectrum \citep{fNL_SS1}.

Applying the general Fisher matrix formalism presented in Sec. \ref{sec:FisherPk} to the forecast for $f_{\rm NL}$, we simply use Eq.~(\ref{eq:Fisher}) but replace $b$ with $b+\Delta b(k)$. In our analysis we set $p=1$ for the LRGs and ELGs and $p=1.6$ for the clustering quasars \citep{fNL1}. We report constraints for the standard local ansatz using $\delta_c=1.686$. We also forecast constraints on $\alpha$ and $\mathcal{A}_{\rm NL}$, expanding around their fiducial values.

Note that the multi-tracer method can provide large improvements on $f_{\rm NL}$ constraints \citep{multi-tracer1,multi-tracer2}, as it measures bias ratios well, and these depend on $f_{\rm NL}$.

 \begin{figure}
\centering
{\includegraphics[scale=0.25]{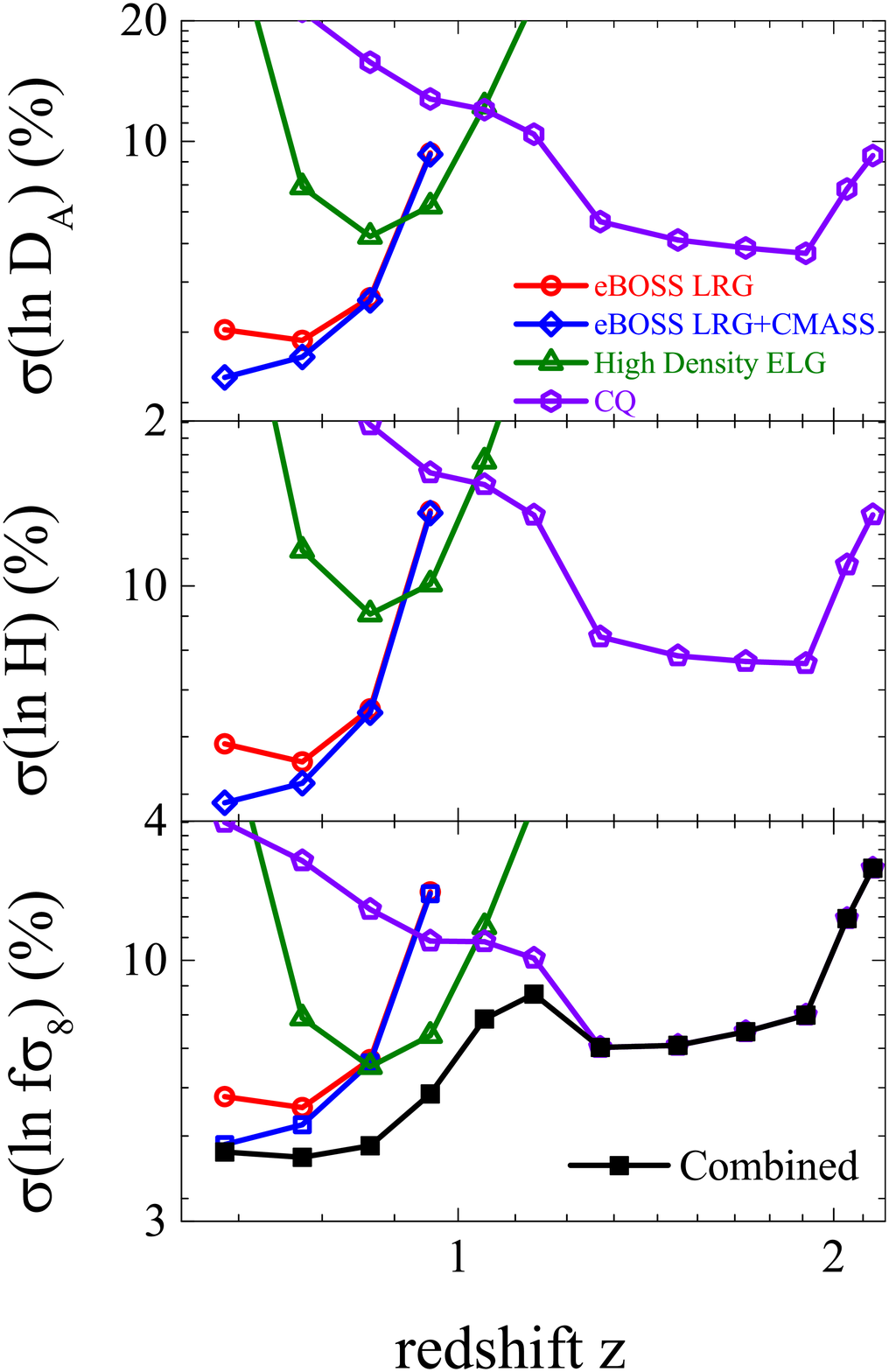}}

\caption{The BAO and RSD constraint using various eBOSS tracers.}
\label{fig:BAORSD}
\end{figure}

 \begin{figure}
\centering
{\includegraphics[scale=0.3]{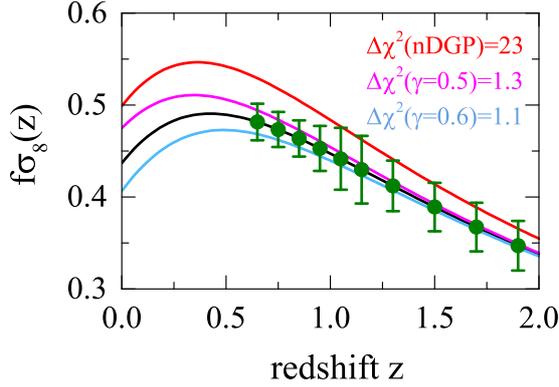}}

\caption{The predicted constraint on $f\sigma_8$ as a function of redshift using combinations of three eBOSS tracers. All models have the same background expansion, giving the same comoving BAO position. The black curve shows the growth in a $\Lambda${CDM} Universe, assuming the Planck best fit model parameters. The red curve shows the nDGP model \citep{DGP} with $\Omega_{r_c} = 0.17$, which corresponds to $r_c H_0 = 1.2$, and $\sigma_8 =0.90$. The magenta and blue curves shows two phenomenological modified gravity models with $\gamma=0.5$ and $\gamma=0.6$ respectively where $f(z) =\Omega_{M}(z)^{\gamma}$ \citep{MGpara7}.}
\label{fig:MG}
\end{figure}
 
\subsection{The Fisher matrix for CMB and WL surveys}
\label{sec:FCMBWL}
Cosmic Microwave Background (CMB) and Weak Lensing (WL) surveys provide highly complementary cosmological information to galaxy redshift surveys. When eBOSS completes its mission in 2020, the full Dark Energy Survey (DES) dataset will be available, in addition to the full Planck data \citep{planck}. Therefore it makes sense to combine the predicted DES, Planck and eBOSS datasets for cosmological forecasts. In this subsection, we briefly describe the formalism, survey specifications and assumptions used to forecast for DES and Planck constraints.  

We assume that the data product of WL experiments are the 2-point angular power spectra $C_{\ell}$, as is the case for the Planck survey. Then the Fisher matrix for parameters $\{p_i,p_j\}$ is \citep{Fisher}, \be F_{ij} =
f_{\rm sky} \sum_{\ell=\ell_{\rm min}}^{\ell_{\rm max}}\frac{2\ell +
1}{2} {\rm Tr}\left( \frac{\partial {\bf C_\ell}}{\partial p_i} {\bf
\tilde{C}_\ell^{-1}}\frac{\partial {\bf C_\ell}}{\partial p_j} {\bf
\tilde{C}_\ell^{-1}} \right) \ , \label{eq:FisherCl} \ee where ${\bf
\tilde{C}_\ell}$ is the observed data covariance matrix with elements
$\tilde{C}^{XY}_\ell$ including noise: \be
\tilde{C}^{XY}_\ell= C^{XY}_\ell+N^{XY}_\ell \ . \label{eq:NoiseAdd}
\ee The quantity $f_{\rm sky}$ is the fraction of sky being surveyed, and the minimum and maximum multipole $\ell_{\rm min}$ and $\ell_{\rm max}$ is set to be, \be \ell_{\rm min} = \pi
/(2{f^{1/2}_{\rm sky}}); \ \ell_{\rm max} = k_{\rm cut} \chi \ee where $\chi$ is the comoving distance from $z=0$ to the redshift slice in which the power spectra are measured, and we set $k_{\rm cut}=0.1 \ {\rm Mpc}^{-1} h$ to avoid using observables in the nonlinear regime.   

DES should ultimately comprise 5000 deg$^2$ of multi-band, optical imaging probing the redshift range $0.1 < z < 1.3$ with a median redshift of $z_0=0.7$ and an approximate
1-$\sigma$ error of $0.05$ in photometric redshift, \ie, $f_{\rm{sky}}=0.13$ and $\sigma(z)=0.05(1+z)$. We take the total galaxy number density distribution to be \citep{galdist}, 
\begin{equation}
N_G(z) \propto z^2 {\rm exp}(-z/z_0)^{2} \ ,
\end{equation} To resolve the radial mode, we subdivde the galaxies into multiple redshift slices and each slice is modelled as, \begin{equation}
N_{G_i}(z) = \frac{1}{2} N_G(z)\left[
  \mathrm{erfc}\biggl( \frac{z_{i-1}-z} {\sqrt{2} \sigma(z)}\biggr)
- \mathrm{erfc}\biggl( \frac{z_i-z}{\sqrt{2} \sigma(z)}\biggr) \right],
\label{eq:erfc}
\end{equation} where erfc is the complementary error function.

We use the WL shear power spectra of DES, and subdivide the total galaxies into four redshift slices. We model the noise power spectra to be \be
N^{\kappa_i \kappa_j}_\ell = \delta_{ij} \frac{\gamma^2_{\rm rms}}{n_j} \label{eq:Noise}
\ee where $\gamma_{\rm rms}$ is the root mean square shear from the intrinsic ellipticity of the galaxies, and $n_j$ is the total number in the $j$th redshift slice. We assume a projected angular density of galaxies $N_G=10$ gal$/$arcmin$^2$, and $\gamma_{\rm rms}=0.18+0.042 \ z$ for DES. Cosmological forecasts using this specification of DES include \citet{LP05,MGCAMB1,MGPCA1}.

We choose the sensitivity of the Planck satellite for the CMB forecast and use the temperature and polarisation angular power spectra. The noise power spectra for the CMB are \citep{LP05},
\ba N^T_{\ell,{\rm c}}&=&(\Delta_T \theta_{\rm FWHM, c})^2 {\rm exp}\left[\frac{\ell(\ell+1)\theta_{\rm FWHM, c})^2}{8~{\rm ln}2}  \right] \nn \\
N^P_{\ell,{\rm c}}&=&(\Delta_P \theta_{\rm FWHM, c})^2 {\rm exp}\left[\frac{\ell(\ell+1)\theta_{\rm FWHM, c})^2}{8~{\rm ln}2}  \right] \ea
where $T$ and $P$ denote the `Temperature' and `Polarisation' respectively, and $\theta_{\rm FWHM, c}$ is the Full width at half maximum (FWHM) of the angular resolution for a given frequency channel c. The combined noise from all channels is then, \ba N^T_{\ell}&=& \left[\sum_{c} \left(N^T_{\ell,{\rm c}}   \right)^{-1} \right]^{-1} \nn \\
N^P_{\ell}&=& \left[\sum_{c} \left(N^P_{\ell,{\rm c}}   \right)^{-1} \right]^{-1}\ea

\subsection{Parametrisations}
\label{sec:para}
The general parametrisation we use is presented in Table \ref{tab:param}, where we list the collection of all the parameters with their physical meaning, and the fiducial value used in the forecast. Note that, however, different subsets of this collection are used in different cases, as detailed in the rest of this subsection. 

\subsubsection{The Parametrisation for the BAO forecast}

As described in Sec. \ref{sec:FisherBAO} and listed as ${\rm \bf P_I}$ of Table \ref{tab:param}, the free parameters for the BAO forecast are ${\rm ln}(D_A/s)(z_i)$ and ${\rm ln}(sH)(z_i)$ in the redshift slice $z_i$. Thus for $N_{z{\rm bin}}$ slices, the total number of BAO parameters are $2N_{z{\rm bin}}$.    

\subsubsection{The Parametrisation for the RSD forecast}

We follow the parametrisation used in \citet{FisherRSD}, namely, for each redshift slice $z_i$, the free parameters for the RSD forecast are ${\rm ln}[f\sigma_8(z_i)]$ and ${\rm ln}[b\sigma_8(z_i)]$. Thus there are $2N_{z{\rm bin}}$ RSD parameters for $N_{z{\rm bin}}$ slices in total. The RSD parameters are listed as ${\rm \bf P_{II}}$ of Table \ref{tab:param} and described in Sec. \ref{sec:FisherRSD}.

\subsubsection{The Parametrisation for the non-Gaussianity forecast}

As described in Sec. \ref{sec:fNL}, the free parameters for the non-Gaussianity forecast for the local model are $f_{\rm NL}$ and $b(z_i,{\rm T}_j)$ where the indices $i,j$ are for the redshift slices and the type of tracer, respectively. So for a redshift survey with $N_{\rm T}$ tracers and $N_{z{\rm bin}}$ redshift slices, the total number of parameters is $N_{\rm T}\times N_{z{\rm bin}}+1$. We will also consider whether this data can constrain departures from the local ansatz, in which case we have an additional parameter, $\alpha$. 

\subsubsection{The Parametrisation for the baseline cosmology}

We use the six-parameter $\Lambda$CDM model, also dubbed the `vanilla' model, as the baseline cosmology model. The parameters of this model are listed as ${\rm \bf P_{III}}$ of Table \ref{tab:param}.

\subsubsection{The Parametrisation for the dark energy forecast}
\label{sec:DEpara}

To forecast for the equation-of-state of dark energy, we adopt two different sets of parametrisations,  

\begin{description}
\item[(I)] The Chevallier-Polarski-Linder (CPL) parametrisation \citep{CP,Linder}: \be w(z) = w_0+w_a \frac{z}{1+z}\ee The free parameters are $w_0,w_a$ with the vanilla parameters ${\rm \bf P_{III}}$; 
\item[(II)] Binned $w$: we discretise $w(z)$ into $M+1$ piece-wise constant bins in $z$ allowing the value of $w$ in each bin to be an independent parameter. Since eBOSS will not be able to probe $z>3$, we use $M$ bins linearly seperated in $z$ for $0\leq z \leq 3$ and a single bin for $z>3$. This allows a principle component analysis (PCA) to be undertaken in Sec. \ref{sec:result_DE}. We take $M=20$ and vary these parameters together with the baseline parameters. \end{description}

\subsubsection{The Parametrisation for the modified gravity forecast}
\label{sec:MGpara}

We follow \citet{MGPCA1} to take the most general parametrisation for modified gravity. Working in the Newtonian gauge, the perturbed Friedmann-Robertson-Walker metric to the first order is,
\begin{equation}\label{FRW}
ds^2=-a^2(\eta)[(1+2\Psi(\vec{x},\eta))d\eta^2-(1-2\Phi(\vec{x},\eta))d\vec{x}^2],
\nonumber
\end{equation}
where $\eta$ is the conformal time and $a(\eta)$ the scale factor.
In Fourier space, we write~\citep{Hu:2007pj,BZ},
\ba
\label{eq:MG} k^2\Psi&=&-\mu(k,a) 4\pi G a^2\rho\Delta \nonumber \\
\Phi/\Psi&=&\eta(k,a)
\ea
where $\Delta$ is the comoving matter density perturbation. The functions $\mu$ and $\eta$ parametrise the MG effect: the function $\eta$ describes anisotropic stresses, while $\mu$ quantifies a time- and scale-dependent rescaling of Newton's constant
$G$. In $\Lambda$CDM, $\mu=\eta=1$ since the anisotropic stress due to radiation is negligible in late times. 

Similar to binning $w(z)$, we treat $\mu(k,z)$ and $\eta(k,z)$ as unknown functions and forecast how well we can constrain the eigenmodes of them using PCA. Since they are 2-variable functions in both $k$ and $a$, we have to bin them in the $(k,z)$ plane. We use the same $M+1$ $z$-bins as $w$ (see Sec. \ref{sec:DEpara}) and $N$ $k$-bins ($0 \leq z \leq 30,10^{-5} \leq k \leq 0.2~{\rm h}\,{\rm Mpc}^{-1}$), with each of the $(M+1)\times N$ pixels having independent values of $\mu_{ij}$ and $\eta_{ij}$. We consider $w(z)$ as another unknown function with independent values in each of the 
$M+1$ $z$-bins. We choose $M=N=20$ and have checked that this binning is fine enough to ensure the convergence of the results. We use logarithmic $k$-bins on superhorizon scales and linear $k$-bins on sub-horizon scales, to optimise computational efficiency. As in~\cite{MGCAMB1}, we only consider information from scales well-described by linear perturbation theory, which is only a fraction of the $(k,z)$-volume probed by future surveys. Since the evolution equations contain time-derivatives of $\mu(k,z)$, $\eta(k,z)$ and $w(z)$, we follow~\cite{wPCA2} and \cite{MGPCA1} and use hyperbolic tangent functions to represent steps in these functions in the $z$-direction, while steps in the $k$-direction are left as step functions. 

Similar to the PCA of $w(z)$, the pixilisation of $\mu(k,z)$ and $\eta(k,z)$ is for the later 2-D PCA, as detailed in \citet{MGPCA1,MGPCA2,MGPCA3,PCASKA}. 

\subsubsection{The Parametrisation for the neutrino mass forecast}

To forecast for the neutrino mass constraint, we vary the sum of neutrino masses with the vanilla cosmological model parameters, \ie, $\sum m_{\nu}$ and the ${\rm\bf P_{III}}$ parameters in Table \ref{tab:param}, and take $\sum m_{\nu}=0.06 \ {\rm eV}$ as the fiducial model. 

\begin{figure*}
\centering
{\includegraphics[scale=0.6]{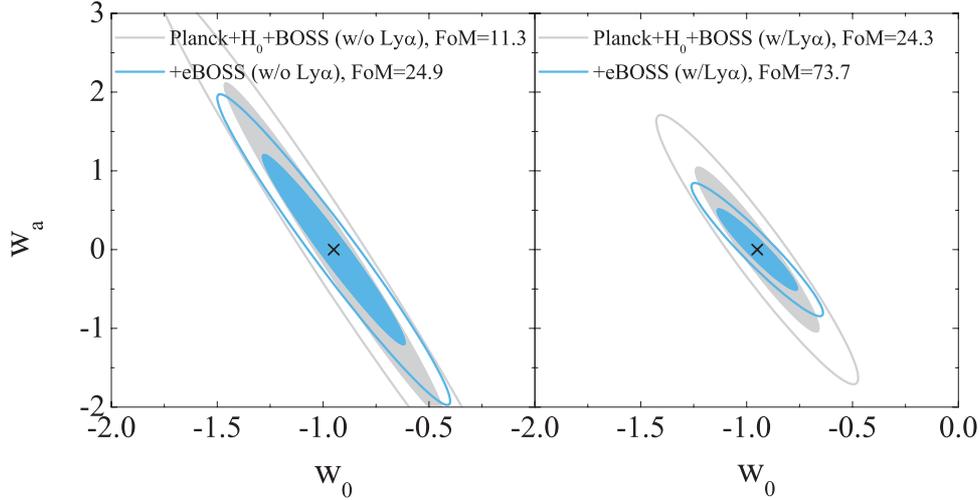}}
\caption{Current constraints on the DETF model for time-varying dark energy compared to projected constraints from eBOSS. We report constraints from the BAO probes, Planck, and $H_0$ from HST observations of SNe Ia. For all measurements, the filled ellipse represents the 68\% confidence interval and the open ellipse represents the 95\% confidence interval.}
\label{fig:FoM}
\end{figure*}

\begin{figure}
\centering
{\includegraphics[scale=0.45]{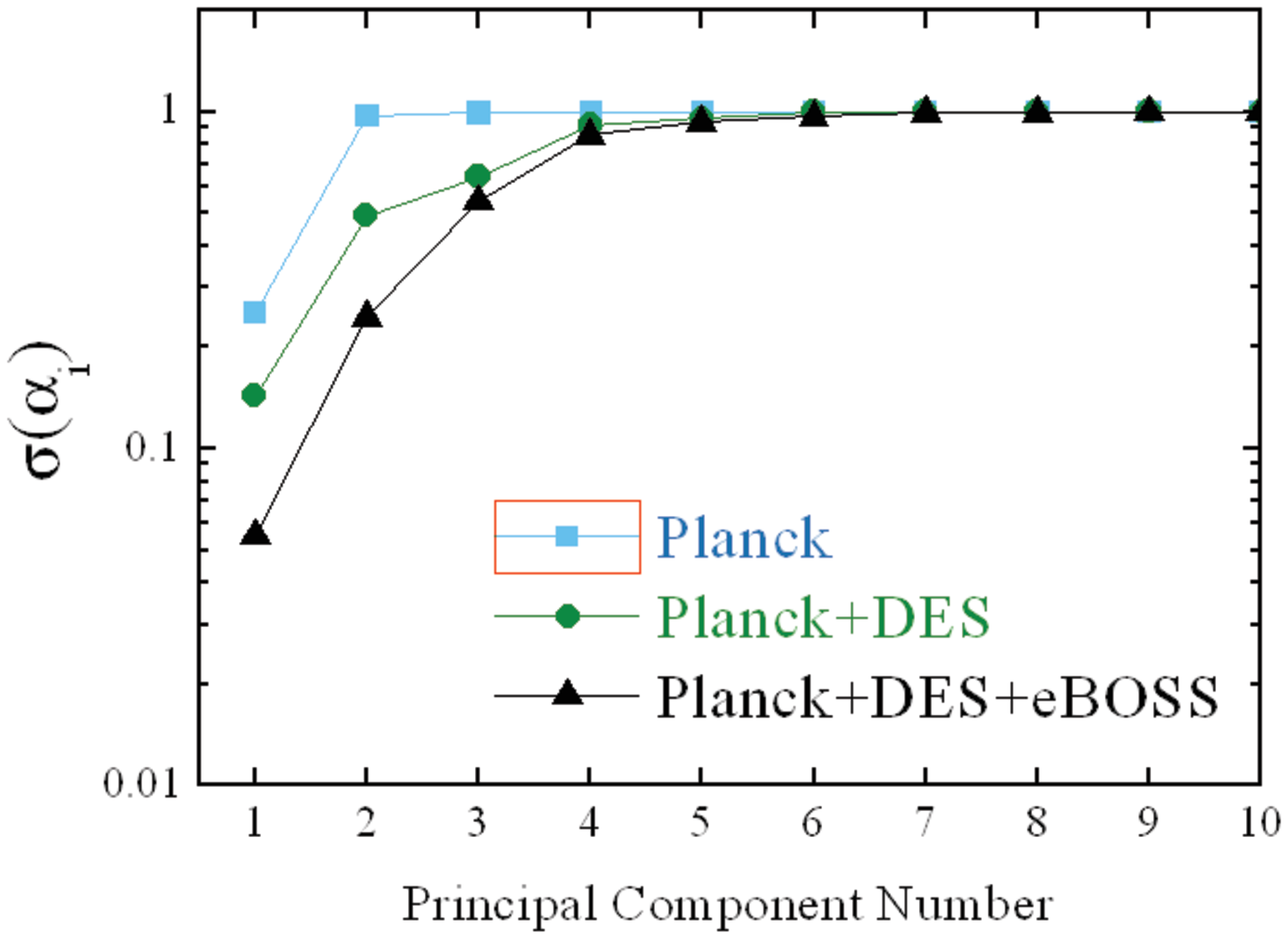}}
{\includegraphics[scale=0.2]{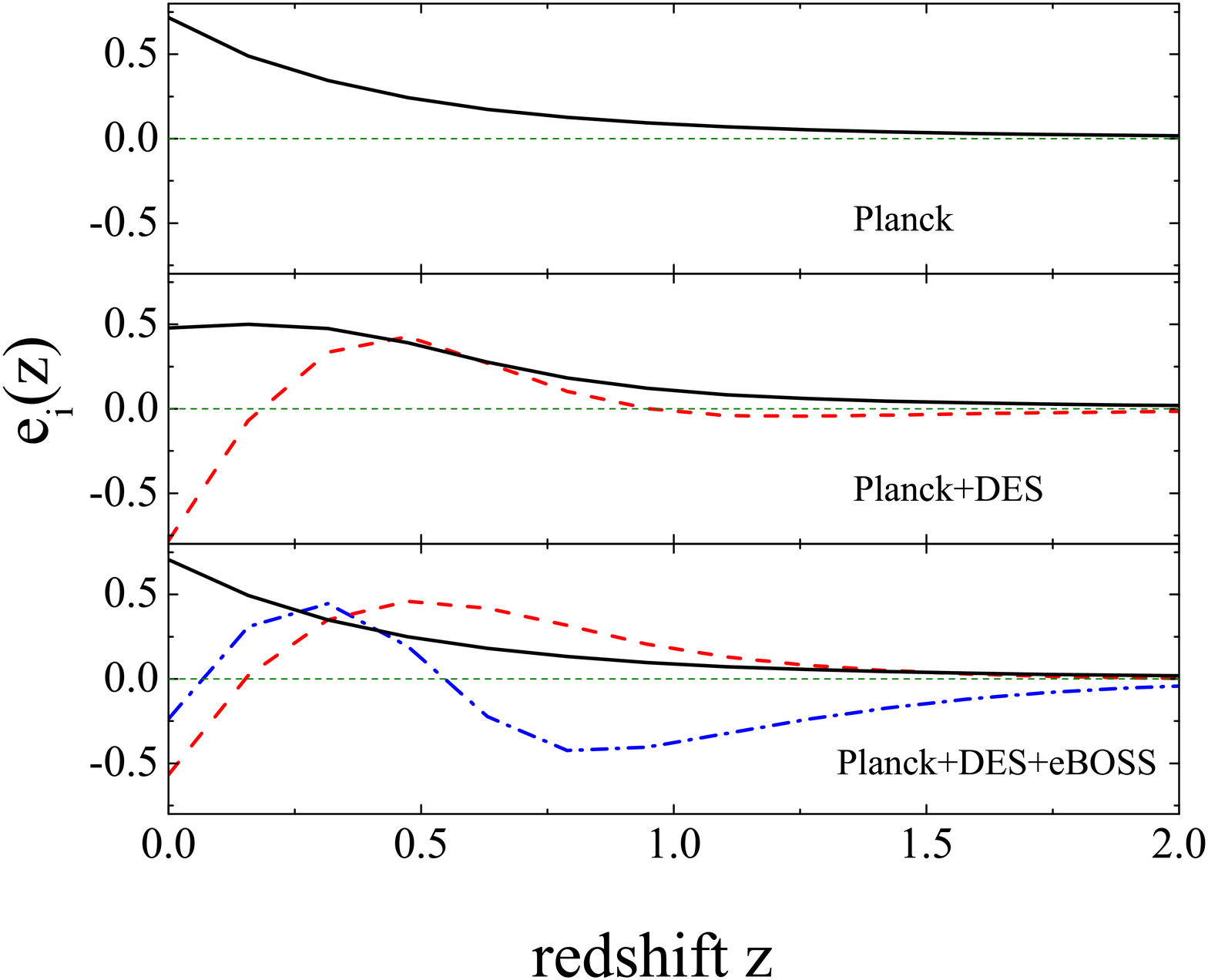}}
\caption{Upper: The forecasted 68\% CL measurement error on $\alpha_i$, the coefficient of the $i$th principal components of $w(z)+1$, namely, $w(z)+1=\sum_i \alpha_i e_i(z)$, using different data combinations illustrated in the legend. A weak prior of $\sigma(w(z))<1$ was assumed. Lower: The best determined eigenvectors (with errors less than $0.5$) of $w(z)$ for different data combinations shown in the legends. The modes are shown, in the order from better constrained to worse, as black solid, red dashed, blue dash-dot, purple dash-dot-dot and brown short dash-dot curves. The short dashed green horizon line shows $e_i(z)=0$. }
\label{fig:wPCA1}
\end{figure}

\section{Results}
\label{sec:result}

In this section, we shall first present the expected precision of BAO and RSD measurements, and then the constraint on general cosmological parameters, including the non-Gaussianity, dark energy and modified gravity parameters and the neutrino mass. 

\subsection{The BAO and RSD forecast}

The primary BAO and RSD forecasts are shown in Tables \ref{tab:BAORSD} and \ref{tab:RSDcombo} and in Figs \ref{fig:BAOLRG} and \ref{fig:BAORSD}. Table \ref{tab:BAORSD} lists the predicted 68\% CL fractional uncertainty on the BAO and RSD parameters, including the angular diameter distance $D_A(z)$, the Hubble parameter $H(z)$, the combined distance $D_V\equiv[c z (1+z)^2 D_A(z)^2 H^{-1}(z)]^{1/3}$, $f(z)\sigma_8(z)$ and $b(z)\sigma_8(z)$ using various tracers in multiple redshift slices. For each tracer, we also list the forecast result at the effective redshift (in bold), and we find that eBOSS LRGs ($0.6<z<1.0$, combined with the BOSS LRGs at $z>0.6$), ELGs ($0.6<z<1.2$) and CQs ($0.6<z<2.2$) can achieve the 1.0\%, 2.3\% and 1.6\% precision respectively for the $D_V$ measurement. Using the same samples, the constraint on $f\sigma_8$ is expected to be 2.5\%, 3.4\% and 2.8\% respectively.

Fig \ref{fig:BAOLRG} shows the forecasted BAO distance using eBOSS LRGs, in comparison with the BOSS measurement. The solid curves show the $\Lambda$CDM prediction, and the upper and lower limits of the bands correspond to the CPL model with $w_0=-1.5, \ w_a=1.0$ and $w_0=-0.5, \ w_a=-1$, respectively. As shown, the eBOSS LRGs sample effectively extends the redshift range of BOSS to higher redshifts with a comparable precision on the distance measurement, namely, 1\% sensitivity on $D_V$ over $0.6<z<1.0$.  

Table \ref{tab:RSDcombo} shows the constraint on $f\sigma_8$ using the three different combined eBOSS samples, depending on the ELGs target selection option, namely, Combined I: LRGs+Fisher ELGs+CQs;  Combined II: LRGs+low density ELGs+CQs; Combined III: LRGs+high density ELGs+CQs. The cross-correlation of the power spectra are included in the overlapping region of different tracers.

Fig \ref{fig:BAORSD} shows the fractional constraint on $D_A$, $H$ and $f\sigma_8$ for individual tracers and for the three combined samples. From this figure, we can tell that, 

\begin{itemize}
\item The CMASS LRGs sample at $z>0.6$ is very helpful for both distance and RSD measurements at $0.6<z<0.7$, \eg, it improves the eBOSS LRGs constraint on $D_A$ from 3.4\% to 2.6\%, and improves the $f\sigma_8$ constraint from 5.3\% to 4.3\%; 
\item The constraints from the ELGs samples are generally weaker than those using the LRGs samples, namely, the uncertainty is roughly larger by a factor of $2$, and $1.4$ for the distance and RSD measurement respectively; 
\item We find that the three different ELGs target selection options yield similar results, especially when combined with LRGs and CQ. The high density selection option has the highest $z_{\rm eff}$, being $0.863$. Thus in the following cosmological forecasts, we choose to use this option for the ELGs to form a combined eBOSS sample, dubbed `Combined III'; 
\item For the distance measurement using the combined eBOSS sample (Combined III), we expect to have 1\%, 2\% and 1.6\% sensitivity on $D_V$ at the effective redshifts of $0.71, 0.86$ and $1.37$ using the LRGs, ELGs and CQs samples respectively;
\item For the $f\sigma_8$ measurement, the LRGs, ELGs and CQs provide a 2.5\%, 3.4\% and 2.8\% precision at the effective redshifts of $0.71, 0.86$ and $1.37$ respectively. Considering narrow slices, the combined sample will allow between 4\% and 15\% precision to be obtained in redshift slices that are between 0.1 and 0.2 thick (See Table \ref{tab:RSDcombo} for details).    
\end{itemize}

Fig \ref{fig:MG} shows the predicted $f\sigma_8$ measurement errors using the combined eBOSS samples, together with the theoretical predictions for the $\Lambda$CDM, nDGP with $r_c H_0=1.2$, and two phenomenological MG models parametrised by the growth index $\gamma$, \ie, $f(z)=\Omega(z)^\gamma$ \citep{MGpara7}. The models with $\gamma=0.5$ and $\gamma=0.6$ are shown. As shown, eBOSS alone can distinguish the $\gamma=0.5$ and $\gamma=0.6$ models from the $\Lambda$CDM model, and can rule out the nDGP model at a significance of 4.8 $\sigma$. 

\subsection{Primordial non-Gaussianity}

The forecast result for $f_{\rm NL}$ (the local ansatz) is given in Table \ref{tab:fNL}, where we show the 68\% CL predicted error on $f_{\rm NL}$ using different tracers individually and three combinations of eBOSS data, depending on the target selection plan for the ELG. As shown, \be \sigma(f_{\rm NL})\sim15 \ ({\rm bias \  float}); \  \sigma(f_{\rm NL})\sim10.5 \ ({\rm bias  \ fixed})\ee

In addition, we consider the more general parametrisation given in Eq.(\ref{eq:generalNGbias}). Since the data is insufficient to constrain the scale dependence of the non-Gaussian bias as a free parameter, we choose fiducial values for $\alpha$ and report constraints on the amplitude and scale dependence. For example, choosing $\mathcal{A}_{\rm NL}=5$ at a pivot scale of $k=0.1\,{\rm Mpc}^{-1}$ (and fixing the Gaussian bias of all tracers), the Combined I dataset yields $\sigma(\mathcal{A}_{\rm NL})=18$ and $\sigma(\alpha)=2.6$ at 65\% C.L.. This result is not very constraining, but it will be interesting to combine the eventual eBOSS LRGs sample with the full BOSS sample (see the results in \citealt{fNL_SS1}) to obtain a tighter constraint.

\begin{table}
\begin{center}
\begin{tabular}{cccc}
\hline\hline
      Sample    & redshift   &  $\sigma(f_{\rm NL})$ & $\sigma(f_{\rm NL})$    \\
                      &                &            (bias float)        &   (bias fixed) \\  
       \hline
      CMASS LRGs &$0.6<z<1.0$  &     37.99   &      24.22       \\  
      eBOSS LRGs &$0.6<z<1.0$ &     23.73    &    15.62        \\   
 CMASS+eBOSS LRGs  &$0.6<z<1.0$ &     22.63 &  14.52           \\   
   Fisher ELGs         &$0.6<z<1.2$ &     94.75    &    56.94        \\    
    Low Density ELGs         &$0.6<z<1.2$  &     87.98  &    52.41            \\    
   High Density ELGs         &$0.6<z<1.2$ &     92.61   &   53.78           \\    
    Clustering Quasars     &$0.6<z<2.2$ &   20.56     & 15.74                  \\   \hline
     Combined I      &$0.6<z<2.2$ &   15.03   &10.50               \\  
     Combined II    &$0.6<z<2.2$ &   15.01     &10.47          \\            
     Combined III   &$0.6<z<2.2$&   15.03     &10.48           \\ 
     \hline\hline  
\end{tabular}
 \end{center}
\caption{Forecasted results of $f_{\rm NL}$ using different tracers individually and three combinations of eBOSS data, depending on the target selection plan for the ELG. The results with and without marginalisation over the bias factor are shown.}    
\label{tab:fNL}
\end{table} 

\subsection{Other cosmological parameters}

In this subsection, we make predictions of how sensitive the full eBOSS galaxy power spectrum will be to cosmological parameters, when combined with external datasets including CMB and weak lensing. We form an eBOSS dataset by combining the LRGs (with the BOSS LRGs at $z>0.6$), high-density ELGs, clustering quasars with all the cross-correlations included. We also include BAO measurements at $z\sim2.4$ using Ly-$\alpha$ forest from BOSS \citep{BAOlya1,BAOlya2,BAOlya3} and eBOSS \citep{Overview}, at 2\% and 1.2\% precision respectively \footnote{There are other possibilities to measure the BAO signal using the triply ionized carbon (C IV) as a tracer in the eBOSS survey, as discussed in \citet{CIV}.}. We refer to this combined data as `eBOSS' in the following forecasts, unless specifically mentioned otherwise. We use {\tt MGCAMB} \citep{MGCAMB1,MGCAMB2} \footnote{Available at \url{http://icosmology.info/MGCAMB.html}}, which is a modified version of {\tt CAMB} \citep{CAMB} to calculate the observables and use {\tt CosmoFish} \footnote{For more information about the CosmoFish package, check \url{http://icosmology.info/cosmofish.html}} for the Fisher matrix calculation. We include the dark energy perturbation following the prescription in \citet{DEP}.  

\subsubsection{Dark energy EoS}
\label{sec:result_DE}

The result for the CPL parametrisation is presented in Fig \ref{fig:FoM}, where the 68 and 95\% CL contour plots of $\{w_0, \ w_a\}$ are shown. The grey and blue contours illustrate the result for BOSS and BOSS+eBOSS respectively \footnote{When combining BOSS with eBOSS galaxies, we take BOSS galaxies in the redshift range of $z<0.6$ to avoid double counting.} combined with the full Planck data and the $H_0$ measurement, and the left and right panels show the prediction without and with the Lyman-$\alpha$ forest data combined. As shown in the legend, the Figure of Merit (FoM), which is inversely proportional to the area of the contours, can be improved by a factor of 3.0 (2.2) when the eBOSS data is combined with (without) the Lyman-$\alpha$ forest. The Lyman-$\alpha$ forest, which provides BAO measurement at high redshift, is highly complementary to the BAO measurement using BOSS/eBOSS galaxies at lower-redshifts since the former can help to break the degeneracy between dark energy parameters and $\Omega_M$, and the latter provides more direct constraint since dark energy dominates at low redshifts, \ie, $z\lesssim1$.

The factor-of-$3$ improvement on the FoM motivated us to explore more details of $w(z)$ using eBOSS by going beyond the CPL parametrisation. As described in (II) of Sec. \ref{sec:DEpara}, we forecast the binned $w(z)$, obtained the Fisher matrix of the $w$ bins ${\bf F_{w}}$ with all other cosmological parameters marginalised over, and perform a PCA on ${\bf F_{w}}$ to determine the eigenmodes that can be well constrained, \ie, \be   {\bf F_w} = {\bf W}^T {\bf \Lambda} {\bf W} \ee where the $i$th rows of the decomposition matrix ${\bf W}$, $e_i(z)$ is the $i$th eigenvector of $w(z)$ and the $(i,i)$ element of the diagonal matrix ${\bf \Lambda}$ stores the corresponding eigenvalue, $\lambda_i$. This enables an orthonormal decomposition of arbitrary $w(z)$, \ie, \be 1+w(z) = \sum_{i=1}^{N} \alpha_i e_i(z), \ \lambda_i=\sigma(\alpha_i)^{-2} \ee We refer the readers to \citet{wPCA1,wPCA2,wPCA3} for more details of PCA for $w(z)$. 

Fig \ref{fig:wPCA1} shows the PCA result using Planck, Planck+DES, Planck+DES+eBOSS data respectively. As shown, the uncertainty of the best constrained eigenmode using eBOSS data (combined with Planck and DES) is $\sim0.05$, and there are three modes which can be measured with uncertainty below $0.5$.

\subsubsection{Modified gravity}

\begin{figure}
\centering
{\includegraphics[scale=0.4]{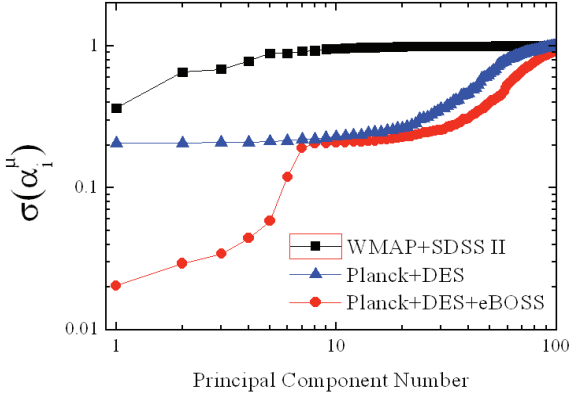}}
\caption{The forecasted 68\% CL error on the coefficients of the principal components of $\mu(k,z)$ for different data combinations shown in the legend.}
\label{fig:MGPCA_eval}
\end{figure}

As described in Sec. \ref{sec:MGpara}, we bin the functions $\mu(k,z)$ and $\eta(k,z)$, and obtain a Fisher matrix for all the bins with all other cosmological parameters marginalised over. Then as for the PCA procedure used for $w(z)$, we perform a PCA on the $\mu$ and $\eta$ functions (we marginalise over all the $\eta$ bins when performing the PCA on $\mu$ and {\it vice versa}). 

It is clear from Eq (\ref{eq:MG}) that $\mu$ determines the growth of structure via the modified Poisson equation, so it can be constrained by redshift surveys like eBOSS. On the other hand, $\eta$ affects the lensing potential thus it could be probed by the CMB and WL surveys instead. Since the purpose of this paper is to highlight the cosmological potential of eBOSS, we show the PCA result of $\mu$ only in Fig \ref{fig:MGPCA_eval}. eBOSS significantly augments the Planck+DES constraint on the $\mu$ modes. The uncertainty on the best-constrained mode is reduced by a factor of $10$, and eBOSS helps to constrain 5 modes to a precision better than 10\%. 

\subsection{Neutrino Mass}

\begin{figure}
\centering
{\includegraphics[scale=0.3]{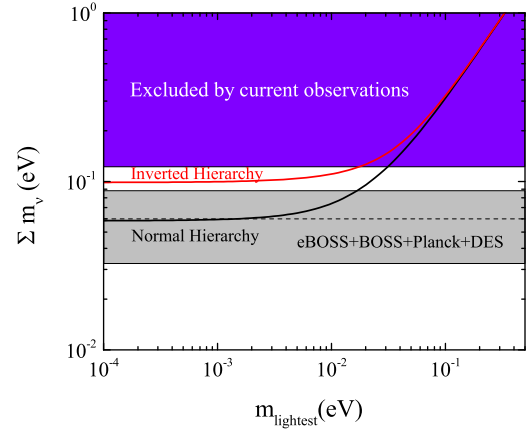}}
\caption{The neutrino mass constraint. The purple shaded region is excluded by the current observations, and the grey shaded band is the expected 68\% CL uncertainty using the full eBOSS survey combined with BOSS, DES and Planck. The black and red curves illustrate the theoretical prediction for the normal and inverted neutrino mass hierarchies.}
\label{fig:neutrino}
\end{figure}

The total neutrino mass as a function of the mass of the lightest species is plotted in Fig \ref{fig:neutrino} to illustrate the normal and inverted mass hierarchy, which are degenerate at the high mass end but in principle distinguishable at the low-mass end by cosmological probes. 

BOSS, combined with other current surveys, has put an upper limit on the neutrino mass of $\sum m_{\nu}<0.12$ eV (95\% CL) \citep{mnu2015,mnulya2}, which is shown by the purple shaded region in Fig \ref{fig:neutrino}. This is not enough to distinguish between NH and IH. Assuming the fiducial value of the total neutrino mass to be $0.06$ eV and using eBOSS combined with BOSS, DES and Planck, we predict the error on the neutrino mass to be \be \sigma\left(\Sigma m_{\nu}\right)=0.03 {\rm eV}\ee This is sufficient to break the degeneracy between the NH and IH scenarios, as shown in Fig \ref{fig:neutrino}, \eg, a measurement of $\sigma\left(\Sigma m_{\nu}\right)<0.06\pm0.03$ would rule out the IH at $1\sigma$ level.
  
\section{Conclusions and discussions}
\label{sec:conclusion}

As the successor of the BOSS survey, the eBOSS survey of the SDSS-IV project is the largest current spectroscopic survey in the world. eBOSS will map the Universe in the redshift range ${0.6 < z < 2.2}$ using multiple tracers and thereby improve our knowledge of the nature of dark energy, test models of gravity, constrain the initial conditions of the Universe, and measure the sum of the mass of neutrinos.

In this work, we have investigated the ability of the eBOSS survey to make BAO distance and RSD growth-rate measurements, and explored the potential of eBOSS for the studies of dark energy, modified gravity, the primordial non-Gaussianity, and the neutrino mass.   

We find that eBOSS will provide strong BAO and RSD measurements in the redshift range of $0.6<z<2.2$ using tracers of the LRGs, ELGs and CQs, namely, the eBOSS LRGs (combined with the BOSS LRGs at $z>0.6$), ELGs and CQs will reach 1\%, 2\% and 1.6\% sensitivity on the BAO distance $D_V$ measurement at effective redshifts of $0.71, 0.86$ and $1.37$ respectively. The RSD effect quantified by $f\sigma_8$, will be measured at a sensitivity of 2.5\%, 3.4\% and 2.8\% by these tracers at the same effective redshifts respectively. The recent work of \citet{ZPW} provides a promising approach to optimise distance-redshift measurements in the BAO. Introducing a small number of redshift weights are demonstrated on a toy model to preserve nearly all of the BAO information at different redshifts. Such an optimisation will be particularly effective for surveys like eBOSS which spans a wide range of redshift. A similar $z$-weighting technique is also likely to be efficient in improving growth rate measurements from the RSD signal \citep{z-weight-RSD1,z-weight-RSD2}. 

The exquisite BAO and RSD measurements that eBOSS will provide are key for dark energy and gravity studies. Choosing a CPL parametrisation for the equation-of-state of dark energy, eBOSS can improve the FoM of dark energy by a factor of 3, with respect to the current BOSS measurement. A more general PCA study of $w(z)$ reveals that eBOSS, combined with DES and Planck, will be able to measure 3 eigenmodes of $w(z)$ with 5\% precision. For modified gravity, a general PCA study finds that eBOSS can significantly improve the constraint on the eigenmodes of $\mu$, the effective Newton's constant, enhancing the DES + Planck constraint. Specifically, eBOSS can improve the constraint on the best-determined mode by a factor of 10, and make it possible to measure 5 modes better than the 10\% accuracy. 

We find that eBOSS alone can determine $f_{\rm NL}$, the parameter quantifying primordial non-Gaussianity, to a precision of $\sigma(f_{\rm NL})=10$ in the optimistic case in which the bias can be well determined separately. When combined with DES and Planck, eBOSS can weigh neutrinos to a precision of $\sigma(\sum m_{\nu})=0.03$ eV, which makes it possible to determine the neutrino mass hierarchy.

\section*{Acknowledgements}

We thank Nicolas Busca, Andreu Font-Ribera and Matthew Pieri for insightful discussions on the Ly-$\alpha$ measurement and forecast. 

GBZ and YW are supported by the Strategic Priority
Research Program "The Emergence of Cosmological
Structures" of the Chinese Academy of Sciences Grant
No. XDB09000000. GBZ is supported by the 1000 Young Talents program in China. YW is supported by the NSFC grant No. 11403034. AH and LP are supported by NSERC. KK is supported by the STFC through the consolidated grant ST/K00090/1, and the European Research Council through grant 646702 (CosTesGrav). GR is supported by the National Research Foundation of Korea (NRF) through NRF-SGER 2014055950.

Funding for the Sloan Digital Sky Survey IV has been provided by
the Alfred P. Sloan Foundation, the U.S. Department of Energy Office of
Science, and the Participating Institutions. SDSS-IV acknowledges
support and resources from the Center for High-Performance Computing at
the University of Utah. The SDSS web site is \url{www.sdss.org}.

SDSS-IV is managed by the Astrophysical Research Consortium for the 
Participating Institutions of the SDSS Collaboration including the 
Brazilian Participation Group, the Carnegie Institution for Science, 
Carnegie Mellon University, the Chilean Participation Group, the French Participation Group, Harvard-Smithsonian Center for Astrophysics, 
Instituto de Astrof\'isica de Canarias, The Johns Hopkins University, 
Kavli Institute for the Physics and Mathematics of the Universe (IPMU) / 
University of Tokyo, Lawrence Berkeley National Laboratory, 
Leibniz Institut f\"ur Astrophysik Potsdam (AIP),  
Max-Planck-Institut f\"ur Astronomie (MPIA Heidelberg), 
Max-Planck-Institut f\"ur Astrophysik (MPA Garching), 
Max-Planck-Institut f\"ur Extraterrestrische Physik (MPE), 
National Astronomical Observatory of China, New Mexico State University, 
New York University, University of Notre Dame, 
Observat\'ario Nacional / MCTI, The Ohio State University, 
Pennsylvania State University, Shanghai Astronomical Observatory, 
United Kingdom Participation Group,
Universidad Nacional Aut\'onoma de M\'exico, University of Arizona, 
University of Colorado Boulder, University of Oxford, University of Portsmouth, 
University of Utah, University of Virginia, University of Washington, University of Wisconsin, 
Vanderbilt University, and Yale University.

\bibliographystyle{mnras}
\bibliography{eBOSS}

\label{lastpage}

\newpage

\appendix

\section{The explicit Fisher matrix for the double-tracer case}

The Fisher matrix for a given $\bf k$ mode, $\mathcal{F}_{ij}(k,\mu)$ in Eq (\ref{eq:F_matrix}), can be calculated explicitly as, \ba\label{eq:F_double_sum} \mathcal{F}_{ij}(k,\mu)=\mathcal{F}_{ij}^{\rm AA}(k,\mu)+\mathcal{F}_{ij}^{\rm BB}(k,\mu)+\mathcal{F}_{ij}^{\rm X}(k,\mu)  \ea
where the Fisher matrices for tracers A, B and their cross-correlation X are,
\ba \label{eq:F_double}\mathcal{F}_{ij}^{\rm AA}(k,\mu)& =& \frac{1}{2}D_{i}^{\rm A}D_{j}^{\rm A} R_{V}^{\rm AA} \nn \\
\mathcal{F}_{ij}^{\rm BB}(k,\mu) &=& \frac{1}{2} D_{i}^{\rm B}D_{j}^{\rm B} R_{V}^{\rm BB} \nn \\
\mathcal{F}_{ij}^{\rm X}(k,\mu) &=&  \ D_{i}^{\rm X}D_{j}^{\rm X} R_{V}^{\rm XX} \nn \\
&& -   \left(D_{i}^{\rm X}D_{j}^{\rm A} + D_{i}^{\rm A}D_{j}^{\rm X}\right) R_{V}^{\rm XA}  \nn \\
&& -   \left(D_{i}^{\rm X}D_{j}^{\rm B} + D_{i}^{\rm B}D_{j}^{\rm X}\right) R_{V}^{\rm XB} \nn \\
&& +\frac{1}{2} \left(D_{i}^{\rm A}D_{j}^{\rm B} + D_{i}^{\rm B}D_{j}^{\rm A}\right) R_{V}^{\rm AB}
 \ea

The derivative for the parameter $p_i$ for tracer T is defined as \be D_i^{\rm T}= \frac{\partial \ {\rm ln}  {P_{{\rm T}}}}{\partial p_i} \ee where ${\rm T}=\{{\rm A, B, X} \}$. The power spectra and the effective volumes are,

\ba  &&R^{\rm AA}_{V}= \left[\frac{n_{\rm A}P_{\rm A}\left(1+n_{\rm B}P_{\rm B}\right)}{ \left(1+n_{\rm A}P_{\rm A}\right)\left(1 + n_{\rm B}P_{\rm B}\right)-n_{\rm A}n_{\rm B} P_{\rm X}^2} \right]^2  \nn\\ 
&&R^{\rm BB}_{V}= \left[\frac{n_{\rm B}P_{\rm B}\left(1+n_{\rm A}P_{\rm A}\right)}{ \left(1+n_{\rm A}P_{\rm A}\right)\left(1 + n_{\rm B}P_{\rm B}\right)-n_{\rm A}n_{\rm B} P_{\rm X}^2 } \right]^2  \nn\\ 
&&R^{\rm XX}_{V}= \frac{ n_{\rm A}n_{\rm B}\left[  \left(1+n_{\rm A}P_{\rm A}\right)\left(1+n_{\rm B}P_{\rm B}\right)  + n_{\rm A}n_{\rm B} P_{\rm X}^2\right] }{ \left[\left(1+n_{\rm A}P_{\rm A}\right)\left(1 + n_{\rm B}P_{\rm B}\right)-n_{\rm A}n_{\rm B} P_{\rm X}^2\right]^2 } P_{\rm X}^2  \nn\\  
&&R^{\rm XA}_{V}= \frac{ n_{\rm A}^2 n_{\rm B}  \left(1+n_{\rm B}P_{\rm B}\right)    }{[\left(1+n_{\rm A}P_{\rm A}\right)\left(1 + n_{\rm B}P_{\rm B}\right)-n_{\rm A}n_{\rm B} P_{\rm X}^2]^2}  P_{\rm X}^2 P_{\rm A}  \nn\\  
&&R^{\rm XB}_{V}= \frac{ n_{\rm A} n_{\rm B}^2  \left(1+n_{\rm A}P_{\rm A}\right)  }{ [\left(1+n_{\rm A}P_{\rm A}\right)\left(1 + n_{\rm B}P_{\rm B}\right)-n_{\rm A}n_{\rm B} P_{\rm X}^2]^2}  P_{\rm X}^2 P_{\rm B}  \nn\\  
&&R^{\rm AB}_{V}= \frac{   n_{\rm A}^2 n_{\rm B}^2 P_{\rm A}  P_{\rm B}  P_{\rm X}^2  }{ [\left(1+n_{\rm A}P_{\rm A}\right)\left(1 + n_{\rm B}P_{\rm B}\right)-n_{\rm A}n_{\rm B} P_{\rm X}^2]^2} \ea

Let us consider several special cases, 

\begin{itemize}
\item The single-tracer limit: $n_{\rm B}=P_{\rm X} \rightarrow 0$. In this case, only $\mathcal{F}_{ij}^{\rm AA}(k,\mu)$ in Eq (\ref{eq:F_double}) is nonzero and it can be easily shown that it recovers the single-tracer result in Eq (\ref{eq:F_single}), namely, \be\label{eq:F_2to1}  \mathcal{F}_{ij}(k,\mu)= \frac{1}{2}D_{i}^{\rm A}D_{j}^{\rm A}  \left(\frac{n_{\rm A}P_{\rm A}}{ 1+n_{\rm A}P_{\rm A}} \right)^2 \ee   
\item The two-independent-tracer limit: $P_{\rm X}\rightarrow0$. Only $R^{\rm AA}_{V}$ and $R^{\rm BB}_{V}$ are nonzero thus the total Fisher matrix is the sum of $\mathcal{F}_{ij}^{\rm AA}(k,\mu)$ and $\mathcal{F}_{ij}^{\rm BB}(k,\mu)$. It is easily shown that, \ba \mathcal{F}_{ij}(k,\mu)=\frac{1}{2} D_{i}^{\rm A}D_{j}^{\rm A}  \left(\frac{n_{\rm A}P_{\rm A}}{ 1+n_{\rm A}P_{\rm A}} \right)^2 \nonumber \\  + \frac{1}{2}D_{i}^{\rm B}D_{j}^{\rm B}  \left(\frac{n_{\rm B}P_{\rm B}}{ 1+n_{\rm B}P_{\rm B}} \right)^2 \ea This is simply the result for two independent tracers.
\item The split-tracer limit: $P_{\rm A}=P_{\rm B}=P_{\rm X}$, $n_{\rm A}=n_{\rm B} \rightarrow n_{\rm A}/2$. This basically splits the same kind of tracer, say, tracer A, into two identical parts, so that the power spectra perfectly correlate with each other, and each subsample has one half of the total number of galaxies. In this case, all terms survive, and after some calculation, the final result turns out to be the same as the single tracer case, \ie, Eq (\ref{eq:F_2to1}). This makes sense intuitively because two halves make one. A generalisation also holds, say, if the same sample is arbitrarily split into two subsamples, the total Fisher matrix (with all the cross-correlation terms included) is the same as the original one without splitting.   
\end{itemize}

\end{document}